%% file: cbf.tex
\documentclass[acmsmall,authorversion]{acmart}

\usepackage{booktabs} 
\usepackage{enumitem} 

\input{includes}
\setcopyright{acmlicensed}
\acmYear{2024} 
\acmDOI{10.1145/3675371}

\acmJournal{PACMCGIT}
\acmVolume{7} 
\acmNumber{3} 
\acmArticle{} 
\acmMonth{7}

\begin{document}
\title{
    Concurrent Binary Trees for Large-Scale Game Components
}

\author{Anis Benyoub}
\orcid{0009-0007-4609-1728}
\affiliation{%
 \institution{Intel Corporation}
 \country{France}
}

\author{Jonathan Dupuy}
\orcid{0000-0002-4447-3147}
\affiliation{%
 \institution{Intel Corporation}
 \country{France}
}

\input{section-teaser}

\begin{abstract}
A concurrent binary tree (CBT) is a GPU-friendly data-structure 
suitable for the generation of bisection based terrain 
tessellations, i.e., adaptive triangulations over square domains. 
In this paper, we expand the benefits of this data-structure in 
two respects. First, we show how to bring bisection based tessellations to 
arbitrary polygon meshes rather than just squares. 
Our approach consists of mapping a triangular 
subdivision primitive, which we refer to as a bisector, 
to each halfedge of the input mesh. 
These bisectors can then be subdivided adaptively to produce conforming 
triangulations solely based on halfedge operators.
Second, we alleviate a limitation that restricted the 
triangulations to low subdivision levels.
We do so by using the CBT as a memory pool manager rather than 
an implicit encoding of the triangulation as done originally. 
By using a CBT in this way, we concurrently allocate and/or 
release bisectors during adaptive subdivision using shared GPU memory.
We demonstrate the benefits of our improvements by rendering
planetary scale geometry out of very coarse meshes. 
Performance-wise, our triangulation method evaluates in less than 0.2ms on 
console-level hardware.
\end{abstract}
%
%


%
%


\maketitle


\section{Introduction}

\paragraph{Motivation}
Modern game engines provide support for large-scale 
components such as terrains, water systems, or planetary 
globes as shown in Figure~\ref{fig_teaser}.
Due to their intrinsic size and resolution, traditional 
level-of-detail approaches like Nanite~\cite{Karis2021} 
are not applicable as the amounts of geometry involved
are simply too large\footnote{as an example, storing the fully subdivided mesh of 
Figure~\ref{fig_teaser} would require exa-Bytes of data, which is orders 
away from what a modern SSD can store}. Instead, these large-scale components 
rely on more ad-hoc solutions that tend to fail in some respect.
For instance: 
\begin{itemize}
    \item[-] clipmaps~\cite{Losasso2004,Asirvatham2005} coupled with 
    static meshes for large vistas~\cite{UEWaterSystem2023} are hard to 
    balance because they depend on the viewing altitude,
    \item[-] projected grids~\cite{Neyret2002} work well for animated surfaces but 
    produce artifacts on static geometry during camera flythroughs,
    \item[-] adaptive grids such as quadtrees are either 
    implemented on the CPU~\cite{Etienne2023} or limited in 
    resolution~\cite{Dupuy2020,Deliot2021}.
\end{itemize}
In this work, we describe an adaptive meshing scheme that 
is free from the aforementioned limitations and suitable 
for managing level-of-detail for the geometry of 
these specialized components.

\paragraph{Contributions and Outline}
Our meshing scheme computes adaptive triangulations on the GPU progressively through time.
As such, it produces optimized amounts of geometry at any time and from ground to space. 
In terms of contributions, it represents an 
improvement of the concurrent binary tree (CBT) method~\cite{Dupuy2020} in two 
distinct aspects, which we detail in the remainder of this article:
\begin{itemize}
    \item In Section~\ref{sec_subdivision}, we describe an algorithm 
    similar to ROAM~\cite{duchaineau1997roaming} that produces adaptive triangulations 
    for halfedge meshes rather than just squares as done in the original CBT paper.
    \item In Section~\ref{sec_cbt}, we describe a parallel implementation of this sequential 
    algorithm by using a CBT as a memory manager, i.e., in a different way than suggested in 
    the original CBT paper.
    \item In Section~\ref{sec_results}, we validate our approach through various rendering results and detailed performance measurements.
\end{itemize}
Thanks to the way we improve, re-use and combine the ROAM and CBT algorithms, we 
render planetary surfaces with centimetric detail entirely on mid-range GPUs 
and using a single representation. 
Our source code is available online: \href{https://github.com/AnisB/large_cbt}{https://github.com/AnisB/large\_cbt}.

\section{Adaptive Triangulation Algorithm}
\label{sec_subdivision}

In this section, we describe the adaptive triangulation algorithm
that we will implement in parallel in Section~\ref{sec_cbt}. 
We start by providing some background and  
position our algorithm with respect to related work 
(Section~\ref{sec_subdivision_background}). 
Next, we describe how we initialize our algorithm using halfedge 
operators (Section~\ref{sec_subdivision_initialization}). 
Finally, we provide each operation necessary to compute 
adaptive triangulations progressively through time 
(Section~\ref{sec_subdivision_algorithm}).

\subsection{Background}
\label{sec_subdivision_background}

\paragraph*{Halfedge Meshes as Input}
Our input solely consists of a halfedge mesh. 
A halfedge mesh~\cite{Halfedge1985,Halfedge1999,PolygonMeshProcessing2010} 
represents a polygon mesh as a set of vertex points 
and halfedges where each halfedge is characterized by its operators 
\TwinID, \NextID, \PrevID, \VertID, \EdgeID, and \FaceID. 
Our implementation relies on a generalization of 
directed edges~\cite{Directed1998,DupuyVanhoey2021}, which is 
illustrated in Figure~\ref{fig_halfedge} along with all the 
aforementioned halfedge operators.

\begin{figure*}
        \input{./figures/figure-halfedge.tex} 
    \caption{\label{fig_halfedge}
        Halfedge mesh representation used for our bisection based subdivision scheme. 
        The blue and green halfedges are highlighted as illustrative examples of a regular and border halfedge, respectively. 
        \vspace{2em}
    }
\end{figure*}

\begin{figure*}
    \begin{center}
        \input{./figures/figure-bisection-overview.tex} 
    \end{center}
    \vspace*{-0.0cm}
    \caption{\label{fig_overview} 
    Overview of our triangulation method.
    \vspace{2em}
    }
\end{figure*}

\begin{figure*}
    \begin{center}
        \input{./figures/figure-compatibility-chain.tex} 
    \end{center}
    \vspace*{-0.0cm}
    \caption{\label{fig_compatibility_chain} 
    Refinement over the compatibility chain of bisector $b_{14}^1$ 
    (highlighted in orange).
    }
\end{figure*}

\paragraph*{Bisection Based Triangulation}
Our triangulation algorithm is illustrated in 
Figure~\ref{fig_overview} and consists in bisecting triangles 
along one of their edges.  
Triangle bisection arises frequently in the scientific literature, 
though under different names. Possible names 
include, e.g., bisection refinement~\cite{maubach1995local},
triangle bintrees~\cite{duchaineau1997roaming}, 4-8~refinement~\cite{velho2000semi,velho20014}, 
diamonds~\cite{hwa2004adaptive,weiss2011simplex,weiss2011diamond,yalccin2011gpu},
right-triangulated irregular networks~\cite{lindstrom1996real,evans2001right}, 
hierarchical simplicial meshes~\cite{atalay2007pointerless}, 
newest vertex bisection~\cite{mitchell2016}, and finally 
longest-edge bisection~\cite{rivara1984algorithms,ozturan1996worst,lindstrom2002terrain}. 
The main difficulty with triangle bisection is that 
it is known to be NP-hard to initialize on arbitrary 
meshes~\cite{mitchell2016}. Our contribution in this particular 
area is to show that the halfedges of the input mesh provide 
an intrinsic triangulation that provides a valid initialization.
This has two benefits. First, it results in an extremely 
simple algorithm. Second, it preserves the topology of the 
input mesh (as opposed to existing methods, e.g.,~\cite{velho20014}), making it more 
suitable for animated assets. Aside from our original initialization method, the logic 
of the algorithm as described in this section is very similar to the ROAM 
method~\cite{duchaineau1997roaming}. 

\subsection{Initialization: Root Bisectors as Halfedges}
\label{sec_subdivision_initialization}

We initialize root bisectors at the surface of an input halfedge 
mesh as shown in Figure~\ref{fig_overview}~(a,~b). 
The properties of these root bisectors are straightforward:

\paragraph*{A Bisector for Each Halfedge}
Each halfedge maps to exactly one root bisector. 
For instance, the halfedge $h_7$ in 
Figure~\ref{fig_overview}~(a) maps to the root bisector 
$b_7^0$ in Figure~\ref{fig_overview}~(b). Note that we refer 
to a bisector using the notation $b_j^d$, where $d \geq 0$ 
denotes its subdivision level and $j \geq 0$ its index.

\paragraph*{Root Bisector Vertices}
We retrieve the vertices of each bisectors thanks to the 
halfedge operators \NextID~and \PrevID. For instance, the vertices of the 
root bisector $b_7^0$ in Figure~\ref{fig_overview}~(b) are 
\begin{align*}
    \text{the vertex } v_{0} &:= \VertID(h_7), \\
    \text{the vertex } v_{1} &:= \VertID(\NextID(h_7)) = \VertID(h_8),\\
    \text{and the average } 
    v_{2} &:= \frac{1}{5} \left(\VertID(h_7) + \VertID(h_8) + \cdots + \VertID(h_{11}) \right).
\end{align*}
Algorithm~\ref{alg_RootVertices} provides pseudocode for computing these vertices.

\begin{minipage}[t]{0.45\textwidth}
    \begin{algorithm}[H]
        \scriptsize
        \begin{algorithmic}[1]
            \caption{Root Bisector Vertices}
            \label{alg_RootVertices}
            \Function{RootBisectorVertices}{halfedgeID: integer}
            \State nextID $\gets$ \Call{Next}{halfedgeID}
            \State $v_0$ $\gets$ \Call{Vert}{halfedgeID}
            \State $v_1$ $\gets$ \Call{Vert}{nextID}
            \State $v_2$ $\gets$ $v_0$
            \State $n$ $\gets$ 1
            \State $h$ $\gets$ nextID
            \While{$h$ $\neq$ halfedgeID}
            \State $v_2$ $\gets$ $v_2$ $+$ \Call{Vert}{$h$}
            \State $n$ $\gets$ $n + 1$
            \State $h$ $\gets$ \Call{Next}{$h$}
            \EndWhile
            \State $v_2$ $\gets$ $v_2 / n$
            \State \Return $\left [ v_0,\, v_1,\,v_2\right ]^T$
            \EndFunction
        \end{algorithmic}
    \end{algorithm}
\end{minipage}
\hfill
\begin{minipage}[t]{0.45\textwidth}
    \begin{algorithm}[H]
        \scriptsize
        \begin{algorithmic}[1]
            \caption{Bisector Vertices}
            \label{alg_Vertices}
            \Function{BisectorVertices}{$b_j^d$: bisector}
            \State halfedgeID $\gets$ $j / 2^{d}$
            \Comment root bisector index
            \State $M \gets I$
            \Comment identity init.
            \State $h \gets j$
            \While{$h \neq $ halfedgeID}
                \Comment compute subd. matrix
                \State $b \gets $ \Call{BitwiseAnd}{$h$ , 1}
                \State $M \gets M \times M_b$
                \Comment see Eq.~\eqref{eq_splitting_matrix}
                \State $h \gets h / 2$
            \EndWhile
            \State \Return $M \,\times $ \Call{RootBisectorVertices}{halfedgeID}
            \EndFunction
        \end{algorithmic}
    \end{algorithm}
\end{minipage}

\paragraph*{Neighboring Bisectors}
We retrieve neighboring bisectors thanks to the halfedge operators 
\NextID, \PrevID, and \TwinID. For instance, the bisectors 
adjacent to the bisector $b_7^0$ in Figure~\ref{fig_overview}~(b) 
are 
\begin{align*}
    \NextID(h_7) = h_{8} & \mapsto b^0_{8},\\
    \PrevID(h_7) = h_{11} & \mapsto b^0_{11}, \\
    \TwinID(h_7) = h_1  & \mapsto b^0_{1}.
\end{align*}

\subsection{Progressive Subdivision Operators}
\label{sec_subdivision_algorithm}

In the following paragraphs, we provide two key operations, namely:
\begin{enumerate}
    \item bisector \emph{refinement}, which increases the resolution of the mesh, and
    \item bisector \emph{decimation}, which conversely decreases the resolution of the mesh.
\end{enumerate}
A key property of these operations is that they guarantee conforming topologies at any
given time when applied locally (as opposed to globally). 
Furthermore, they allow for "incremental updates" of an existing triangulation 
(typically the one from the previous frame) into a slightly more appropriate one 
in the spirit of ROAM~\cite{duchaineau1997roaming}. This is done by iterating 
over existing bisectors and deciding whether to refine, 
decimate or leave each one of them as is. The key benefit of incremental updates 
is that they provide an alternative to computing entire triangulations in either 
bottom-up or top-down fashion. This is particularly important for large-scale game 
components as they require large subdivision depths.

\paragraph{Notation}
Refining a bisector $b_j^d$ splits it into two new ones. We denote these 
new bisectors as 
\begin{alignat*}{2}
    b_j^d 
    &\mapsto b_{2j}^{d+1} \qquad&& \text{(first child)}
    \\
    &\mapsto b_{2j+1}^{d+1} \qquad&& \text{(second child)}.
\end{alignat*}
In Figure~\ref{fig_compatibility_chain} for instance, the bisector 
$b_{14}^1$ shown in inset~(b)
splits into the bisectors $b_{28}^2$ and $b_{29}^2$ shown in inset~(c). 
The reason why we index the children in this way is because we can retrieve their
vertices straightforwardly through its subdivision matrix as we show next.

\paragraph{Bisector Vertices via its Subdivision Matrix} 
The bisector refinement rule consists in splitting the bisector into 
two new ones according to the following refinement matrices 
(note the difference between the first two rows): 
\begin{equation}
    \label{eq_splitting_matrix}
    M_0 = 
    \begin{bmatrix}
        1 & 0 & 0\\
        0 & 0 & 1\\
        \frac{1}{2} & \frac{1}{2} & 0\\
    \end{bmatrix}
    , \quad \textrm{and} \quad
    M_1 = 
    \begin{bmatrix}
        0 & 0 & 1\\
        0 & 1 & 0\\
        \frac{1}{2} & \frac{1}{2} & 0\\
    \end{bmatrix}.
\end{equation}
Based on the way we index our bisectors, we retrieve the 
subdivision matrix associated with any bisector 
as shown in lines~(3--9) of Algorithm~\ref{alg_Vertices}. 
In order to get the vertices we simply multiply this matrix 
by the vertices of the root bisector as shown in 
line~(10) of Algorithm~\ref{alg_Vertices}.

\paragraph{Bisector Refinement}
We implement the newest vertex bisection algorithm~\cite{mitchell2016}, 
which we provide in Algorithm~\ref{alg_Refine}. 
This algorithm ensures that the triangulations produced by the refinement 
of any bisector results in a conforming triangulation. This is 
done by propagating the refinement operation through a so-called 
compatibility chain, which is unique and determined through the 
recursive calls in Algorithm~\ref{alg_Refine}. We illustrate a 
practical example of such a propagation in 
Figure~\ref{fig_compatibility_chain} for the bisection of 
the bisector $b_{14}^1$ shown in orange. In order to make this 
propagation work, each bisector must have pointers to its direct neighbors, 
and we discuss next how we implement this.

\begin{minipage}[t]{0.45\textwidth}

\begin{algorithm}[H]
    \scriptsize
    \begin{algorithmic}[1]
        \caption{Adaptive Refinement}
        \label{alg_Refine}
        \Procedure{Refine}{$b_j$: bisector}
        \State $b_k$ $\gets$ \Call{Twin}{$b_j$}
        \If{$b_k$ $\neq$ null}
            \If{\Call{Twin}{$b_k$} $\neq$ $b_j$}
            \State \Call{Refine}{$b_k$} 
            \EndIf
            \State \Call{Split}{$b_j$, $b_k$}
            \Comment non-boundary
        \Else
            \State \Call{Split}{$b_j$}
            \Comment boundary
        \EndIf
        \EndProcedure
    \end{algorithmic}
\end{algorithm}
\end{minipage}
\hfill
\begin{minipage}[t]{0.45\textwidth}

\begin{algorithm}[H]
    \scriptsize
    \begin{algorithmic}[1]
        \caption{Conservative Decimation}
        \label{alg_Decimation}
        \Procedure{Decimate}{$b_{j_1}$: bisector}
        \State bitValue $\gets$ \Call{BitwiseAnd}{$j_1$, 1}
        \State $b_{j_2}$ $\gets$ bitValue ? \Call{Prev}{$b_{j_1}$} : \Call{Next}{$b_{j_1}$}
        \State $b_{j_3}$ $\gets$ bitValue ? \Call{Next}{$b_{j_1}$} : \Call{Prev}{$b_{j_1}$}
        
        \If{$\lfloor \frac{j_1}{2} \rfloor$ $=$ $\lfloor \frac{j_2}{2} \rfloor$}
            \If{$b_{j_3}$ $=$ null}
                \State \Call{Merge}{$b_{j_1}$, $b_{j_2}$} \Comment boundary
            \ElsIf{$\lfloor \log_2 {j_1} \rfloor$ $=$ $\lfloor \log_2 {j_3} \rfloor$}
                \State $b_{j_4}$ $\gets$ bitValue ? \Call{Next}{$b_{j_3}$} : \Call{Prev}{$b_{j_3}$}
                \If{$\lfloor \frac{j_3}{2} \rfloor$ $=$ $\lfloor \frac{j_4}{2} \rfloor$}
                    \State \Call{Merge}{$b_{j_1}$, $b_{j_2}$, $b_{j_3}$, $b_{j_4}$} 
                    \Comment non-boundary
                \EndIf
            \EndIf
        \EndIf
        \EndProcedure
    \end{algorithmic}
\end{algorithm}
\end{minipage}

\begin{minipage}[t]{0.45\textwidth}
\begin{algorithm}[H]
    \scriptsize
    \begin{algorithmic}[1]
        \caption{Pointer Refinement}
        \label{alg_NeighborRefine}
        \Procedure{RefinePointers}{$b_j^d$}
        \State next $\gets$ \Call{Next}{$b_j^d$}
        \State prev $\gets$ \Call{Prev}{$b_j^d$} 
        \If{\Call{Prev}{next} $=$ $b_j^d$}
            \State \Call{Prev}{next} $\gets b_{2j+1}^{d+1}$ 
        \Else
            \State \Call{Twin}{next} $\gets b_{2j+1}^{d+1}$
        \EndIf
        \If{\Call{Next}{prev} $=$ $b_j^d$}
            \State \Call{Next}{prev} $\gets b_{2j}^{d+1}$ 
        \Else
            \State \Call{Twin}{prev} $\gets b_{2j}^{d+1}$
        \EndIf
        \EndProcedure
    \end{algorithmic}
\end{algorithm}
\end{minipage}
\hfill
\begin{minipage}[t]{0.45\textwidth}
\begin{algorithm}[H]
    \scriptsize
    \begin{algorithmic}[1]
        \caption{Pointer Decimation}
        \label{alg_NeighborDecimate}
        \Procedure{DecimatePointers}{$b_{2j}^{d+1}, b_{2j+1}^{d+1}$}
        \State next $\gets$ \Call{Twin}{$b_{2j+1}^{d+1}$}
        \State prev $\gets$ \Call{Twin}{$b_{2j}^{d+1}$}
        \If{\Call{Prev}{next} $=$ $b_{2j+1}^{d+1}$}
            \State \Call{Prev}{next} $\gets b_{j}^{d}$ 
        \Else
            \State \Call{Twin}{next} $\gets b_{j}^{d}$
        \EndIf
        \If{\Call{Next}{prev} $=$ $b_{2j}^{d+1}$}
            \State \Call{Next}{prev} $\gets b_{j}^{d}$ 
        \Else
            \State \Call{Twin}{prev} $\gets b_{j}^{d}$
        \EndIf
        \EndProcedure
    \end{algorithmic}
\end{algorithm}
\end{minipage}

\begin{center}
    \begin{table}[h!]
        \small
        \begin{tabular}{l|l}
            {children of $b_j^d$}
            &
            {children of $b_k^d$ (if and only if $b_k^d \neq \text{null}$)}
            \\ \hline \\
            $\begin{aligned}
                \NextID(b_j^d)  
                    \mapsto &\,\NextID(b_{2j}^{d+1})  = \;b_{2j+1}^{d+1} \\
                    \mapsto &\,\NextID(b_{2j+1}^{d+1}) = \begin{cases}
                        b_{2k}^{d+1} & \text{if $b_k^d \neq \text{null}$,} \\
                        \textrm{null} & \text{otherwise.}
                    \end{cases}
            \end{aligned}$
            &
            $\begin{aligned}
                \NextID(b_k^d)  
                    \mapsto &\,\NextID(b_{2k}^{d+1}) = b_{2k+1}^{d+1}  \\
                    \mapsto &\,\NextID(b_{2k+1}^{d+1}) = b_{2j}^{d+1}
            \end{aligned}$
            \\ \\
            $\begin{aligned}
                \PrevID(b_j^d)  
                    \mapsto &\,\PrevID(b_{2j}^{d+1}) = \begin{cases}
                        b_{2k+1}^{d+1} & \text{if $b_k^d \neq \text{null}$,} \\
                        \textrm{null} & \text{otherwise.}
                    \end{cases} \\
                    \mapsto &\,\PrevID(b_{2j+1}^{d+1}) = \; b_{2j}^{d+1}
            \end{aligned}$
            &
            $\begin{aligned}
                \PrevID(b_k^d)  
                    \mapsto &\,\PrevID(b_{2k}^{d+1}) = b_{2j+1}^{d+1}   \\
                    \mapsto &\,\PrevID(b_{2k+1}^{d+1}) = b_{2k}^{d+1}
            \end{aligned}$
            \\ \\
            $\begin{aligned}
                \TwinID(b_j^d)  
                    \mapsto &\,\TwinID(b_{2j}^{d+1}) = \PrevID(b_j^d)
                    \\
                    \mapsto &\,\TwinID(b_{2j+1}^{d+1}) = \NextID(b_j^d)
            \end{aligned}$
            &
            $\begin{aligned}
                \TwinID(b_k^d)  
                    \mapsto &\,\TwinID(b_{2k}^{d+1}) = \PrevID(b_k^d)   \\
                    \mapsto &\,\TwinID(b_{2k+1}^{d+1}) = \NextID(b_k^d)
            \end{aligned}$
            \\
        \end{tabular}
        \vspace{1em}
    \caption{Refinement rules for bisector's neighborhood operators.}
    \label{tab_Neighbors}
    \end{table}
\end{center}

\paragraph{Neighbor Refinement Rule} Bisector refinement operates either: 
\begin{enumerate}
    \item in isolation at a boundary as shown in Figure~\ref{fig_compatibility_chain}~(b),
    \item in pairs of same subdivision level as shown in Figure~\ref{fig_compatibility_chain}~(c).
\end{enumerate}
Let $b_j^d$ and $b_k^d = \TwinID(b_j^d)$ denote the bisectors 
forming such a pair. If $b_j^d$ is located at a boundary, we have 
$b_k^d = \text{null}$. The neighbors of the children 
of $b_j^d$ and $b_k^d$ are given by the rules compiled in Table~\ref{tab_Neighbors}.
Note that when a bisector is refined, its neighbors must also be updated to reference 
the newly created bisectors according to the rules of Algorithm~\ref{alg_NeighborRefine}.

\paragraph{Bisector Decimation} 
Decimation is the reverse operation of refinement and we could simply implement it as such. 
In practice though, we rely on a more conservative approach, which avoids the ambiguous 
case where we decimate entire compatibility chains that carry bisectors under refinement. 
Our approach thus only halves the number of bisectors in the two following configurations:
\begin{enumerate}
    \item we decimate four bisectors of same subdivision depth forming a cycle via their $\NextID$ and 
    $\PrevID$ operators into two coarser ones; in Figure~\ref{fig_compatibility_chain}
    for instance, we decimate the four bisectors $b_{28}^2$, $b_{29}^2$, $b_{46}^2$ and $b_{47}^2$ 
    shown in inset~(c) into $b_{14}^1$ and $b_{23}^1$ shown in inset~(b).
    \item we decimate two border bisectors of same subdivision depth having their 
    $\TwinID$ point to the null bisector (because of the border) into a single one; 
    in Figure~\ref{fig_compatibility_chain} for 
    instance, we decimate the bisectors $b_{22}^1$ and $b_{23}^1$ shown in inset~(b) into $b_{11}^0$ 
    shown in inset~(a).
\end{enumerate}
The resulting logic is shown in Algorithm~\ref{alg_Decimation}.
Note that when a bisector is decimated, its neighbors must also be updated to reference 
the newly created bisector accoring to the rules of Algorithm~\ref{alg_NeighborDecimate}.

\section{GPU Implementation using a CBT}
\label{sec_cbt}

We wish to parallelize the triangulation method from 
Section~\ref{sec_subdivision} over the bisectors already present in the triangulation. 
As such, each bisector should be able to concurrently refine 
(see Algorithm~\ref{alg_Refine}), decimate (see Algorithm~\ref{alg_Decimation}), 
or leave itself untouched. In addition, each bisector should concurrently update 
neighbor pointer information (see Algorithms~\ref{alg_NeighborRefine},~\ref{alg_NeighborDecimate}). 
In this section, we show how to use a concurrent binary tree (CBT) to fulfill these requirements. 
We start by recalling details about the CBT data-structure 
(Section~\ref{sec_cbt_background}). We then show how we use CBTs as 
a memory pool manager and thread scheduler (Section~\ref{sec_cbt_manager}). 
Finally, we provide implementation details (Section~\ref{sec_cbt_triangulation}).

\subsection{Background}
\label{sec_cbt_background}

\paragraph{Positioning}
CBTs were introduced along with a set of algorithms to compute binary subdivisions over 
square domains~\cite{Dupuy2020}. As such, we could have re-used these algorithms for 
the bisector-based scheme from Section~\ref{sec_subdivision}. Unfortunately though, the 
original implementation links the maximum subdivision depth of the binary subdivision to 
that of the CBT. In practice, this restricts the algorithm to subdivisions that are way below 
what is required for planetary scale rendering (note that we further discuss this issue in 
Section~\ref{sec_results}). So rather than re-using the CBT data-structure 
as is, we modify it so that it can manage a memory pool of bisectors. The memory pool 
approach is especially relevant for our needs as our bisectors require the same amount of data 
independently from their subdivision depth. Hence, aside from the memory layout of the CBT, which we recall 
next, we revisit its usage/implementation entirely as a memory manager.

\paragraph{Concurrent Binary Tree Data-Structure}
Figure~\ref{fig_cbt}~(left) shows an example of a CBT of depth $D = 4$. 
A CBT~\cite{Dupuy2020} is a full binary tree where: 
\begin{itemize}
    \item[-] the leaf nodes store binary values and form a bitfield,
    \item[-] the remaining nodes form a sum-reduction tree over these binary values.
\end{itemize}
Note that, by construction, the root node of a CBT stores the number of bits set to one in 
the bitfield.  

\paragraph{Data-Layout}
We store a CBT as a binary heap, i.e., an array of size $2^{D+1}$ as shown in 
Figure~\ref{fig_cbt}~(right). This leads to the following properties:
\begin{itemize} 
    \item[-] the children of the node indexed by $k\geq1$ are located at indexes $2k$ and $2k+1$, 
    \item[-] the parent of the node indexed by $k\geq1$ is located at index $k/2$.  
\end{itemize}
This layout avoids the need to store explicit pointers to parent and children within each nodes 
as they can be computed analytically.

\begin{figure*}
    \begin{center}
        \input{./figures/figure-cbt.tex} 
    \end{center}
    \vspace*{-0.25cm}
    \caption{
        The CBT's (left) data-structure and (right) contiguous memory layout within an array.
    }
    \label{fig_cbt}
\end{figure*}

\subsection{CBT as a Memory Manager}
\label{sec_cbt_manager}
We now provide all the operations required to use a CBT as a memory pool manager. 
Note that we describe these operations in a generic fashion since they 
are not tied to our triangulation algorithm (we defer specializations for our 
triangulation algorithm to the next subsection).

\paragraph{Initialization}
We allocate a CBT of depth $D\geq 1$ that we use in tandem with a memory pool of capacity $2^D$, 
i.e., an array of $2^D$ constant-sized memory blocks. 
These memory blocks must be sized in advance and cannot be re-sized 
dynamically. In order to track which blocks within the memory pool are available, we pair 
the $i$-th bit from the CBT's bitfield to the $i$-th block in the memory pool. 
We thus have a bit value associated to each memory block. By convention, we set bits to one 
when a block is allocated, and zero when freed/available. Thanks to the CBT's sum-reduction 
tree, we know how many blocks are allocated by reading the root 
node. In addition, the CBT's sum-reduction tree allows scheduling a thread to read from 
a valid memory block using the following binary search algorithms. 

\paragraph{Binary Search Algorithm}
Given a CBT of depth $D$, we retrieve the index of the $i$-th bit 
set to one in $O(D)$ as shown in Algorithm~\ref{alg_Decode}. 
For the CBT of Figure~\ref{fig_cbt} for instance, this algorithm returns:
\begin{itemize}
    \item[-] index 0 for the first bit,
    \item[-] index 3 for the second bit,
    \item[-] index 10 for the third bit,
    \item[-] etc.
\end{itemize}
Note that, by duality, it is also possible to retrieve the $i$-th bit set to 
zero with the same complexity, see Algorithm~\ref{alg_Encode}. This dual 
algorithm is important to track available memory blocks.

\begin{minipage}[t]{0.45\textwidth}
\begin{algorithm}[H]
    \scriptsize
    \begin{algorithmic}[1]
        \caption{Find the $i$-th bit set to one}
        \label{alg_Decode}
        \Function{OneToBitID}{index: int}
            \State bitID $\gets 1$ 
            \While{bitID < $2^D$}
            \State bitID $\gets$ 2 $\times$ bitID
            \If {index $\geq$ CBT[bitID]}
                \State index $\gets$ index $-$ CBT[bitID]
                \State bitID $\gets$ bitID $+$ 1
            \EndIf
            \EndWhile
            \State \Return bitID - $2^D$
        \EndFunction
    \end{algorithmic}
\end{algorithm}
\end{minipage}
\hfill
\begin{minipage}[t]{0.45\textwidth}
\begin{algorithm}[H]
    \scriptsize
    \begin{algorithmic}[1]
        \caption{Find the $i$-th bit set to zero}
        \label{alg_Encode}
        \Function{ZeroToBitID}{index: int}
            \State bitID $\gets 1$ 
            \State $c \gets 2^{D-1}$ 
            \While{bitID < $2^D$}
            \State bitID $\gets$ 2 $\times$ bitID
            \If {index $\geq$ ($c - $CBT[bitID])}
                \State index $\gets$ index $-$ ($c -$CBT[bitID])
                \State bitID $\gets$ bitID $+$ 1
            \EndIf
            \State $c \gets c / 2$ 
            \EndWhile
            \State \Return bitID - $2^D$
        \EndFunction
    \end{algorithmic}
\end{algorithm}
\vspace*{0.75em}
\end{minipage}

\paragraph{Usage}
We update the memory pool in parallel through a two-step approach as shown in Algorithm~\ref{alg_Update}.
During the first step (see lines 2--16), we iterate over each valid memory block and 
allow for allocation and de-allocation operations. We detail this step a bit further in the next 
paragraph. Once this step completed, we simply launch a sum reduction over the CBT's updated bitfield. 
This completes the update process.

\paragraph{Block Iteration and Updates}
We parallelize block iteration by scheduling a thread for each memory block. 
Each thread loads its associated memory block using Algorithm~\ref{alg_Decode}. 
Whenever a thread requests a new memory block, we proceed in the following steps:
\begin{enumerate}
    \item we atomically increment an allocation counter that keeps track of the number of allocations, 
    \item the original value of this counter tells us which 0-valued bit in the CBT's bitfield 
    should be set to one thanks to Algorithm~\ref{alg_Encode},
    \item once this bit is located, we set it to one and write data to the associated memory block.
\end{enumerate}
As for the case when the thread requests to erase its associated memory block, we simply set the 
associated bit to zero within the bitfield. After the sum reduction is computed 
at the end of the update, we will be able to overwrite it safely.

\vspace*{0.75em}
\begin{algorithm}
    \scriptsize
    \begin{algorithmic}[1]
        \caption{Typical memory pool update using a CBT}
        \label{alg_Update}
        \Procedure{ProcessMemory}{cbt, pool}
        \State counter $\gets 0$
        \State blockCount $\gets$ cbt[1]
        \ForAll {threadID $\in$ $[$0, blockCount$)$}
        \State blockID $\gets$ \Call{OneToBitID}{cbt, threadID}
        \State $\cdots$
        \While{mallocRequired}
            \State mallocID $\gets$ \Call{AtomicAdd}{counter, 1}    
            \State newBlockID $\gets$ \Call{ZeroToBitID}{cbt, mallocID}
            \State \Call{SetBit}{cbt, newBlockID, 1}
            \State \Call{SetBlock}{pool, newBlockID, newData}
        \EndWhile
        \State $\cdots$
        \If{freeRequired}
            \State SetBit(cbt, blockID, 0)
        \EndIf        
        \EndFor
        \State \Call{UpdateSumReductionTree}{cbt}
        \EndProcedure
    \end{algorithmic}
\end{algorithm}

\clearpage
\subsection{GPU Triangulation Pipeline}
\label{sec_cbt_triangulation}

We provide here implementation details of our CBT-based GPU triangulation algorithm.
The objective of this section is two-fold. First, to provide the necessary details suitable for 
reproducing our implementation. Second, to convince the reader that our parallel implementation is 
viable.

\begin{table}[h]
    \small
    \begin{center}
{
    \begin{tabular}{l|c|l}
        attribute name          & type              & element size          \\ 
        \hline
        (input) halfedge buffer &            array  & $\times 6$ integer    \\
        (input) vertex buffer   &            array  & $\times 3$ floats     \\
        bisector pool           & $2^D$      array  & $\times 9$ integers   \\
        CBT                     & $2^{D+1}$  array  & $\times 1$ integer    \\
        allocation counter      & scalar            & $\times 1$ integer    \\
        pointer buffer          & $2^D$      array  & $\times 2$ integers
    \end{tabular}
}
    \end{center}
    \caption{Memory requirements for our implementation.}
    \label{tab_memreq}
\end{table}

\paragraph*{Memory Requirements}
We list all our memory requirements in Table~\ref{tab_memreq}.
Given a halfedge mesh such as the one shown in Figure~\ref{fig_halfedge}, 
we allocate a CBT of depth $D$. The depth $D$ must be large enough to hold all the 
bisectors of the mesh, i.e., we must have $D \geq \lceil \log_2(H) \rceil$, where $H \geq 3$ 
denotes the number of halfedges of the input mesh. In Figure~\ref{fig_halfedge} for instance, we 
have $H = 12$ so we must have $D \geq 4$. Note that we typically use \mbox{$D \in [16, 19]$} in 
our implementation. 
This CBT manages a memory pool of capacity $2^D$ bisectors in the spirit of what we described 
in Section~\ref{sec_cbt_manager}. We store the following data for each bisector within the memory pool:
\begin{itemize}
    \item[-] ($\times 3$) integers that act as pointers within the memory pool for the neighbors $\NextID$, 
    $\PrevID$, and $\TwinID$ of each bisector,
    \item[-] ($\times 1$) integer for the bisector index, i.e., the value $j$ that is required to 
    decode the vertices of the bisector as shown in Algorithm~\ref{alg_Vertices}. 
\end{itemize}
Note that the width of these integer is important: the neighbor pointers must be 
sufficiently large to address the entire pool, while the bisector index enforces a 
maximum subdivision depth. In our implementation, we use 32-bit integers for the neighbor pointers 
and a 64-bit integer for the bisector index.
In addition to these two attributes, we also require data to concurrently split and/or merge 
each bisector. 
We store the following additional data for each bisector:
\begin{itemize}
    \item[-] ($\times 1$) integer that encodes a split/merge command as a bit code, which we 
    modify concurrently via atomic bitwise-ORs, 
    \item[-] ($\times 4$) integers that store unused memory pool locations where the bisector can write to. 
\end{itemize}
The former integer allows us to make sure splits and/or merges are only evaluated once per 
bisector. The latter ones are sized based on the fact that we only allow one refinement and/or 
decimation level per bisector during each incremental update. This means that in the worst 
case, a bisector will split into four new ones (hence four integers). Such a case happens whenever 
the bisector belongs to the compatibility chain of two foreign bisectors under refinement. 
In Figure~\ref{fig_compatibility_chain} for instance, splitting bisector  
$b_{14}^1$ results in splitting $b_{11}^0$ into three new ones due to its location in $b_{14}^1$'s 
compatibility chain.
Finally, we further allocate memory for the following:
\begin{itemize}
    \item[-] an atomic counter, i.e., an integer, that we increment for each memory 
    allocation,
    \item[-] a $2^D$ buffer of ($\times 2$) integers per element, which caches the results of 
    Algorithms~\ref{alg_Decode}~and~\ref{alg_Encode}. 
\end{itemize}
We further detail the purpose of each of these attributes in the following paragraphs.

\paragraph*{Initialization}
We initialize the memory pool with $H$ root bisectors, where we recall that $H \geq 3$ denotes 
the number of halfedges of the input mesh. We place these bisectors at the first $H$ 
blocks of the memory pool
and set the first $H$ bit of the CBT's bitfield to one (the rest are set to zero). 
In the memory pool, we uniquely index each bisector as 
\begin{equation*}
    \texttt{bisector.index} = \lceil \log_2 H \rceil + h,
\end{equation*} 
where $h \in [0, H-1]$ denotes the index of the mesh's halfedge. As for the neighbors, we 
simply copy the values of the halfedge buffer 
\begin{align*}
    \texttt{bisector.next} &= \NextID(h) \\
    \texttt{bisector.prev} &= \PrevID(h) \\
    \texttt{bisector.twin} &= \TwinID(h).
\end{align*} 
We then compute the sum-reduction tree of the CBT to complete the initialization.
The rest of the memory is used during our incremental updates, which we describe next.

\begin{figure*}
    \begin{center}
        \input{./figures/figure-pipeline.tex} 
    \end{center}
    \caption{\label{fig_pipeline} 
    GPU update loop for our adaptive triangulation algorithm. 
    Memory, CPU commands and GPU kernels are respectively shown in green, gray and red.
    Each kernel is implicitely followed by a GPU memory barrier.
    }
\end{figure*}

\paragraph*{Incremental Update Pipeline}
We update the triangulation incrementally. Each incremental update consists in iterating 
over each bisector in parallel. In turn, these bisectors can either be refined or decimated 
once. In our experiments, this approach is enough to handle planetary scales in realtime 
as we showcase in Section~\ref{sec_results}. 
Our GPU update routine is illustrated in Figure~\ref{fig_pipeline} and consists of 9 
shader kernels, which we detail below:

\paragraph*{\emph{(1)} \texttt{ResetCounter}} 
We simply set the allocation counter to zero. We use this counter to track the number 
of allocations we need to make to update our triangulation. Note that this kernel requires 
a single thread invocation, while all subsequent ones require one thread per bisector, 
whose number we retrieve from the root element of the CBT.

\paragraph*{\emph{(2)} \texttt{CachePointers}}
We store the result of Algorithm~\ref{alg_Decode} in the first integer of the pointer buffer. 
This allows us to precompute the index buffer of bisectors located in the memory pool, which 
we use at the beginning of every subsequent kernel to load the data associated 
with each bisector (i.e., we use \texttt{bisector = MemoryPool[PointerBuffer[threadID]]}). 
We also store the result of Algorithm~\ref{alg_Encode} 
in the second integer of the pointer buffer. This precomputes the index buffer of the 
available blocks in the memory pool, which we use for allocations.

\paragraph*{\emph{(3)} \texttt{ResetCommands}}
We set the command of each bisector to zero. 

\paragraph*{\emph{(4)} \texttt{GenerateCommands}}
Here we decide whether each bisector should be refined, decimated or kept as is. 
In our applications, we make our decision by decompressing the vertices of the 
bisector via Algorithm~\ref{alg_Vertices} and making sure its screen-space area 
matches a user-defined value. 
We record the decision in the command integer using atomic bitwise-OR operations
(by doing so we issue every possible split and merge command only once). We modify 
the command depending on the decision:
\begin{itemize}[wide]
\item[-] In case of refinement, we mimic Algorithm~\ref{alg_Refine} by scattering a 
split command for the active bisector as well as those that belong to the compatibility chain. 
For the bisectors that belong to the compatibility chain, we need to track which edge 
of the bisector must be bisected. This is trivial to do thanks to the neighbor operators. 
Since there are only 3 neighbors per bisector, we reserve 3-bits for the refinement command, 
each bit corresponding to an edge of the bisector. Note that we condition the execution of 
this algorithm to the amount of memory available in the CBT. To do so, we atomically 
increment the allocation counter by the maximum number of allocations. This number 
depends on the subdivision depth $d$ of the bisector and satisfies 
\begin{equation*}
    \texttt{maxAllocation} = 3  d + 4.
\end{equation*} 
Thus we only execute the refinement algorithm if the CBT has enough available bits 
to allocate such a number. In case of overflow, we atomically decrement the allocation counter 
by the same number and skip refinement for this specific bisector.
\item[-] In case of decimation, we mimic Algorithm~\ref{alg_Decimation} by first making 
sure that the bisector belongs to a configuration that can be merged. If it does, 
we write a merge command to the active bisector. 
Our merge command records the merge configuration (either triangle or quad as 
described in Section~\ref{sec_subdivision_algorithm}) and a flag that tags whether 
the bisector has the smallest index of the configuration. The latter bit is important to 
ensure we allocate for the root bisectors only once later in the pipeline. 
We therefore reserve 3-bits (different from those used for refinement) for the merge 
command as well. Just as in the case of refinement, we condition the execution of our 
merge algorithm to the space left in the CBT. Since decimation results in a maximum of 
two allocations, we increment the allocation counter by two and proceed similarly 
to the refinement step. 
\item[-] Finally, if we decide to keep the bisector as is, we leave the command untouched.
\end{itemize}
At the end of this step, each bisector has a specific command and the allocation counter 
gives us the number of allocations we have to perform in the CBT.

\paragraph*{\emph{(5)} \texttt{ReserveBlocks}}
We write to the 4 integers we allocated per bisector to store unused 
memory pool locations. For each bisector, we read its associated command and determine 
how many allocations it must make in the memory pool. Due to the way 
we generate the commands in the previous kernel, a bisector may request splits 
(at most 4) and/or a merge. When both are required (this happens when a bisector decides 
to merge and belongs to the compatibility chain of another bisector that requests a split), 
we ignore the merge command and only apply splits. We then atomically decrement 
the allocation counter by the number of allocations \texttt{nAlloc} required by the 
bisector, and save the memory locations cached in the pointer buffer in the range
[\texttt{counter}, \texttt{counter + nAlloc} - 1]. 
In the case of a merge, only the bisector tagged with the minimum index performs 
allocations. Such allocations are conditioned to the fact that all 
relevant bisectors must agree to merge together, i.e., their command must exclusively request 
a merge. At the end of this step, each bisector 
has a unique set of memory pool locations where it can safely write new data.

\paragraph*{\emph{(6)} \texttt{FillNewBlocks}}
Here each bisector writes data to the memory blocks it has allocated. Specifically, 
we compute and save the index of each newly created bisector. In addition, we set the 
pointers that inherit from the active bisector according to Table~\ref{tab_Neighbors}.
At the end of this step, all bisectors have a valid index and a partial number of 
valid pointers to their neighbors. We update the remaining pointers in the next step.

\paragraph*{\emph{(7)} \texttt{UpdateNeighbors}}
Each bisector scatters data to its neighbors. This step is necessary for the data produced by the 
active bisector's neighbors to properly reference the data produced by the active bisector. 
Depending on the command, we apply the rules (or compositions of these rules in the case 
where more than 2 splits are required) of Algorithm~\ref{alg_NeighborRefine} and 
Algorithm~\ref{alg_NeighborDecimate}. At the end of this step, all the bisectors have 
valid pointers to their neighbors.

\paragraph*{\emph{(8)} \texttt{UpdateBitfield}}
We update the bitfield of the CBT. Any bisector that has a non-null command must be freed 
so we set their associated bit to zero. Depending on the command, we set up to 4 bits to 
one to save the allocations triggered by the bisector.

\paragraph*{\emph{(9)} \texttt{SumReduction}}
We update the sum-reduction tree of the CBT. After this kernel, we know how many memory 
blocks are used and, conversely, free in the memory pool. We can thus trigger a new 
round of incremental update to further refine and/or decimate our input geometry.

\section{Evaluation and Discussion}
\label{sec_results}
In this section, we validate our method and compare its performances against 
alternative ones.

\subsection{Test Scenes}

\paragraph*{Water Systems}
In Figure~\ref{fig_teaser}, we show a water-system running on top of an 
Earth-sized planet (i.e., roughly 510 million square kilometers) rendered using our triangulation method. 
We also show a wireframe view segmented as eight successive insets; 
notice the scale covered by our triangulation and how it adapts 
to the camera's view frustum to produce equally-sized triangle in 
screen-space. The planet's base mesh is a 
regular dodecahedron, which consists of 12 pentagonal faces thus leading 
to $12 \times 5 = 60$ halfedges and root bisectors. 
We refine this base mesh over camera-visible areas so that each 
triangle occupies roughly 49 pixels on-screen. 
During refinement, we project the triangles' vertices 
on the surface of a sphere, and further displace them using 
four layers of choppy waves~\cite{tessendorf2001}. Each layer is 
synthesized per-frame on the GPU as a \mbox{$256\times256$} half-float
texture using an FFT and displace the vertices both 
horizontally and vertically with respect to their tangent frame. 
Once we have updated the triangulation, we export it into a 
visibility buffer~\cite{Burns2013Visibility} that we render
using the GPU's hardware rasterizer. For shading, we rely on physically-based BRDFs 
illuminated by a precomputed atmosphere~\cite{Bruneton2008}. 
Thanks to the adaptivety of our triangulation, we are able to 
render this particular shot using only 64k triangles. 
We refer the reader to our supplemental video for the animated result.

\paragraph*{Terrain Components}
In Figure~\ref{fig_moon}, we show an elevation-based terrain example
over a Moon-sized surface (i.e., roughly 38 million square kilometers) using NASA's moon data, which is available at the 
following URL: \url{https://svs.gsfc.nasa.gov/cgi-bin/details.cgi?aid=4720}.
Our elevation data consists of a \mbox{$5760\times2880$} 16-bit displacement map coupled with several 
octaves of simplex noise. The base mesh and refinement target is the 
same as the one we described for the ocean system in the previous paragraph. 
We also rely on a visibility buffer for rendering. 
For shading, we use a diffuse BRDF with an albedo that varies 
according to a \mbox{$4096\times2048$} texture map.

\paragraph*{Arbitrary Asset}
For the sake of completeness, we also evaluated our method against 
hardware-accelerated tessellation shaders over a complex mesh. 
In Figure~\ref{fig_tessellation}, we compare our adaptive triangulation 
against that of tessellation shaders for a specific view configuration.

\begin{figure*}
    \begin{center}
        \input{./figures/figure-moon.tex} 
    \end{center}
    \vspace*{-0.5cm}
    \caption{\label{fig_moon} 
    (top) Frames taken from a zoom-in animation over the Moon terrain. 
    (bottom) Aeral wireframe views.
    }
\end{figure*}

\begin{figure*}
    \begin{center}
        \input{./figures/figure-tessellation.tex} 
    \end{center}
    \vspace*{-0.5cm}
    \caption{\label{fig_tessellation} 
    Comparison between (left) tessellation shaders and (right) our triangulation on a complex mesh.
    }
\end{figure*}

\subsection{Performance Measurements}

\begin{table}[h]
    \small
    \begin{center}
{
    \begin{tabular}{c|c|c|c}
        test scene  & Figure~\ref{fig_teaser}   & Figure~\ref{fig_moon} & Figure~\ref{fig_tessellation} \\ 
        ($H$ / $D$) & (60 / 17)                 & (60 / 17)             & (21,399 / 18)                 \\ 
        \hline
        \hline
        update      & 0.1                       &  0.084                & 0.085                         \\
        render      & 1.745                     &  1.863                & 0.1   
    \end{tabular}
}
    \end{center}
    \caption{
        Performances for various geometries on an AMD~6800~XT GPU. 
        Timings are given in milliseconds.
    }
    \label{tab_perf}
\end{table}

\paragraph*{Absolute Timings}
We benchmarked our test scenes on an AMD~6800~XT, which we chose to mimic a console-level 
GPU. Since our method runs entirely on the GPU, the CPU does not play a major role in 
performances. Nevertheless, we mention we use an Intel Core i7-13700K for the sake of 
completeness. We report our performance measurements for each test scene in Table~\ref{tab_perf}. 
The update row measures the time taken to compute an incremental update of our triangulation, 
which comprises all 9 kernels from Figure~\ref{fig_pipeline}. As demonstrated by the reported 
numbers, our incremental update is very fast to evaluate, taking less than 0.1ms per frame.
Note that update performances are similar between Figure~\ref{fig_moon} and 
Figure~\ref{fig_tessellation} despite the CBT being twice as big for the latter scene.
This is due to the way we implement our \texttt{GenerateCommands} kernel, which requires 
evaluating the world-space position of each bisector's vertices; this evaluation is more 
expensive for Figure~\ref{fig_teaser} and Figure~\ref{fig_moon} because we rely on double 
precision (otherwise discretization artifacts appear as shown in Figure~\ref{fig_float}). 
Note that it may be possible to avoid the use of double precision via quantization methods 
but we have not tried this.
As for the rendering row, we sum the timings required to create the visibility buffer 
and shade the pixels in the case of Figure~\ref{fig_teaser} and Figure~\ref{fig_moon}.
For Figure~\ref{fig_tessellation} we simply rely on forward rendering. 

\paragraph*{Memory Consumption}
    Here we provide the memory consumption of our implementation for 
    Figure~\ref{fig_teaser} and Figure~\ref{fig_moon}.
    We rely on a CBT of depth $D=17$ and 32-bit integers everywhere except for the 
    bisector index, which is 64-bit wide. This makes a total of 7 MiB of memory (which 
    can be deduced from Table~\ref{tab_memreq}) to handle planetary-scale triangulations.

\paragraph*{Performances Against the Original CBT Implementation}
As mentioned in Section~\ref{sec_cbt}, the original CBT implementation~\cite{Dupuy2020} 
couples the maximum depth of the bisectors to that of the CBT, which is set to depth $D = 27$. 
This has the following implications/limitations: 
\begin{enumerate}
    \item The root bisectors can be subdivided at most 27 times. While this is enough for 
    moderately-sized terrains, it is way too limited for scenes like those of 
    Figure~\ref{fig_teaser} and Figure~\ref{fig_moon}.
    \item The implementation requires to allocate memory for $2^{27}$ elements despite 
    the fact that most of this memory is not used: in practice the adaptive triangulation 
    produces at most 128k triangles for every viewing conditions we have tried, i.e., over 99\% of the memory is wasted.
    \item A sum reduction must be computed over $2^{27}$ elements and is the main bottleneck 
    of the method. On our test hardware, such a computation runs in 0.4ms, which alone is four 
    times slower than our entire update time as reported in Table~\ref{tab_perf}.
\end{enumerate}
In contrast to the original approach, ours decouples the depth of the CBT to that of the 
subdivision. As a consequence, we reach subdivision depths of up to 64 (as we use 64-bit 
integers to encode bisector indexes; deeper levels can be achieved with larger integers) 
with a memory pool of 128k elements and therefore a CBT of depth $D = 17$. Due to the reduced 
size of the CBT, we also avoid a large sum-reduction compared to the original method. 
Our novel approach thus produces deeper subdivision levels with less memory while running 
faster.

\begin{figure*}
    \begin{center}
        \input{./figures/figure-float.tex} 
    \end{center}
    \vspace*{-0.5cm}
    \caption{\label{fig_float} 
    Impact of double floating point precision on planetary-scale tessellation.
    }
\end{figure*}

\section{Conclusion}
We introduced an adaptive triangulation method for arbitrary 
halfedge meshes that runs entirely on the GPU. Our method 
is primarily geared towards large-scale game components such 
as terrains, water systems and planets, for which it delivers high performances 
on console-level GPUs.

\bibliographystyle{ACM-Reference-Format}
\bibliography{references}

\end{document}

%% file: includes.tex
\usepackage{tikz}
\usetikzlibrary{patterns}
\usetikzlibrary{arrows.meta}
\usetikzlibrary{fadings}
\usepackage{fancyvrb}
\usepackage{soul}
\usepackage{bm}
\usepackage{pgfplots}
\usepackage{pgfplotstable}
\usepgfplotslibrary{patchplots,colorbrewer,groupplots}
\usetikzlibrary{shadows,backgrounds,arrows,positioning,matrix,calc}
\pgfplotsset{compat=newest}
\usepackage[export]{adjustbox}
\usepackage{nicefrac}
\usepackage{gensymb}
\usepackage{graphicx}
\usepackage{gensymb}
\usepackage{color, colortbl}
\usepackage{stmaryrd} 
\usepackage{pifont}
\usepackage[figuresleft]{rotating}

\definecolor{myred}{rgb}{0.86,0.00,0.00}
\definecolor{myredlight}{rgb}{0.97,0.75,0.75}
\definecolor{myredlighter}{rgb}{0.99,0.94,0.94}
\definecolor{myredlighterr}{rgb}{1.0,0.98,0.98}
\definecolor{myblue}{rgb}{0.00,0.20,0.70}
\definecolor{mybluelight}{rgb}{0.75,0.80,0.93}
\definecolor{mybluelighter}{rgb}{0.94,0.95,0.98}
\definecolor{mybluelighterr}{rgb}{0.98,0.99,1.0}
\definecolor{mygreen}{rgb}{0.10,0.50,0.10}
\definecolor{mygreenlight}{rgb}{0.78,0.88,0.78}
\definecolor{mygreenlighter}{rgb}{0.94,0.97,0.94}
\definecolor{mygreenlighterr}{rgb}{0.99,0.99,0.99}
\definecolor{mygrey}{rgb}{0.40,0.40,0.40}
\definecolor{mygreylight}{rgb}{0.85,0.85,0.85}
\definecolor{mygreylighter}{rgb}{0.96,0.96,0.96}
\definecolor{mygreylighterr}{rgb}{0.99,0.99,0.99}
\definecolor{myorange}{rgb}{1.0,0.50,0.00}
\definecolor{myorangelight}{rgb}{1.0,0.87,0.75}
\definecolor{myorangelighter}{rgb}{1.0,0.96,0.93}
\definecolor{myorangelighterr}{rgb}{1.0,0.99,0.98}

\usepackage{algorithm}
\usepackage{algpseudocode}

\usepackage{listings}
\lstdefinelanguage{GLSL}%
{%
	morekeywords={%
		false,FALSE,NULL,true,TRUE,%
		__LINE__,__FILE__,__VERSION__,GL_core_profile,GL_es_profile,GL_compatibility_profile,%
		precision,highp,mediump,lowp,%
		break,case,continue,default,discard,do,else,for,if,return,switch,while,%
		void,bool,int,uint,float,double,vec2,vec3,vec4,dvec2,dvec3,dvec4,bvec2,bvec3,bvec4,ivec2,ivec3,ivec4,uvec2,uvec3,uvec4,mat2,mat3,mat4,mat2x2,mat2x3,mat2x4,mat3x2,mat3x3,mat3x4,mat4x2,mat4x3,mat4x4,dmat2,dmat3,dmat4,dmat2x2,dmat2x3,dmat2x4,dmat3x2,dmat3x3,dmat3x4,dmat4x2,dmat4x3,dmat4x4,sampler1D,sampler2D,sampler3D,image1D,image2D,image3D,samplerCube,imageCube,sampler2DRect,image2DRect,sampler1DArray,sampler2DArray,image1DArray,image2DArray,samplerBuffer,imageBuffer,sampler2DMS,image2DMS,sampler2DMSArray,image2DMSArray,samplerCubeArray,imageCubeArray,sampler1DShadow,sampler2DShadow,sampler2DRectShadow,sampler1DArrayShadow,sampler2DArrayShadow,samplerCubeShadow,samplerCubeArrayShadow,isampler1D,isampler2D,isampler3D,iimage1D,iimage2D,iimage3D,isamplerCube,iimageCube,isampler2DRect,iimage2DRect,isampler1DArray,isampler2DArray,iimage1DArray,iimage2DArray,isamplerBuffer,iimageBuffer,isampler2DMS,iimage2DMS,isampler2DMSArray,iimage2DMSArray,isamplerCubeArray,iimageCubeArray,atomic_uint,usampler1D,usampler2D,usampler3D,uimage1D,uimage2D,uimage3D,usamplerCube,uimageCube,usampler2DRect,uimage2DRect,usampler1DArray,usampler2DArray,uimage1DArray,uimage2DArray,usamplerBuffer,uimageBuffer,usampler2DMS,uimage2DMS,usampler2DMSArray,uimage2DMSArray,usamplerCubeArray,uimageCubeArray,struct,%
		abs,acos,all,any,asin,atan,ceil,clamp,cos,cross,degrees,dFdx,dFdy,distance,dot,equal,exp,exp2,faceforward,floor,fma,fract,ftransform,fwidth,greaterThan,greaterThanEqual,inversesqrt,length,lessThan,lessThanEqual,log,log2,matrixCompMult,max,min,mix,mod,noise1,noise2,noise3,noise4,normalize,not,notEqual,outerProduct,pow,radians,reflect,refract,shadow1D,shadow1DLod,shadow1DProj,shadow1DProjLod,shadow2D,shadow2DLod,shadow2DProj,shadow2DProjLod,sign,sin,smoothstep,sqrt,step,tan,texture1D,texture1DLod,texture1DProj,texture1DProjLod,texture2D,texture2DLod,texture2DProj,texture2DProjLod,texture3D,texture3DLod,texture3DProj,texture3DProjLod,textureCube,textureCubeLod,transpose,%
		rgb
	},
	sensitive=true,%
	morecomment=[s]{/*}{*/},%
	morecomment=[l]//,%
	morestring=[b]",%
	morestring=[b]',%
	moredelim=*[directive]\#,%
	moredirectives={define,defined,elif,else,if,ifdef,endif,line,error,ifndef,include,pragma,undef,warning,extension,version}%
}[keywords,comments,strings,directives]%
\definecolor{lstattrib}{rgb}{0,0.34,0}
\DeclareCaptionFormat{mylst}{\hrule#1#2#3}
\lstnewenvironment{cpp}[1][]{\lstset{language=c++, #1}}
    {}
\lstnewenvironment{glsl}[1][]{\lstset{language=GLSL, #1}}
    {}

\newcommand{\TwinID}{\textsc{Twin}}
\newcommand{\PrevID}{\textsc{Prev}}
\newcommand{\NextID}{\textsc{Next}}
\newcommand{\VertID}{\textsc{Vert}}
\newcommand{\EdgeID}{\textsc{Edge}}
\newcommand{\FaceID}{\textsc{Face}}

%% file: section-teaser.tex
\begin{teaserfigure}
    \begin{center}
        \input{figures/figure-teaser}
    \end{center}
\caption{\label{fig_teaser} 
(top) An Earth-sized planet rendered in real-time using our triangulation method.
(bottom) An alternative wireframe-view of the render as seen from ground to space 
through successive zoomed-in insets.
}
\end{teaserfigure}

%% file: figures/figure-teaser.tex
\begin{tikzpicture}[every node/.style={inner sep=0,outer sep=0}]
    \node[anchor = north west] at (0.0, 5.2) {\includegraphics[fbox, width=0.9215\textwidth]{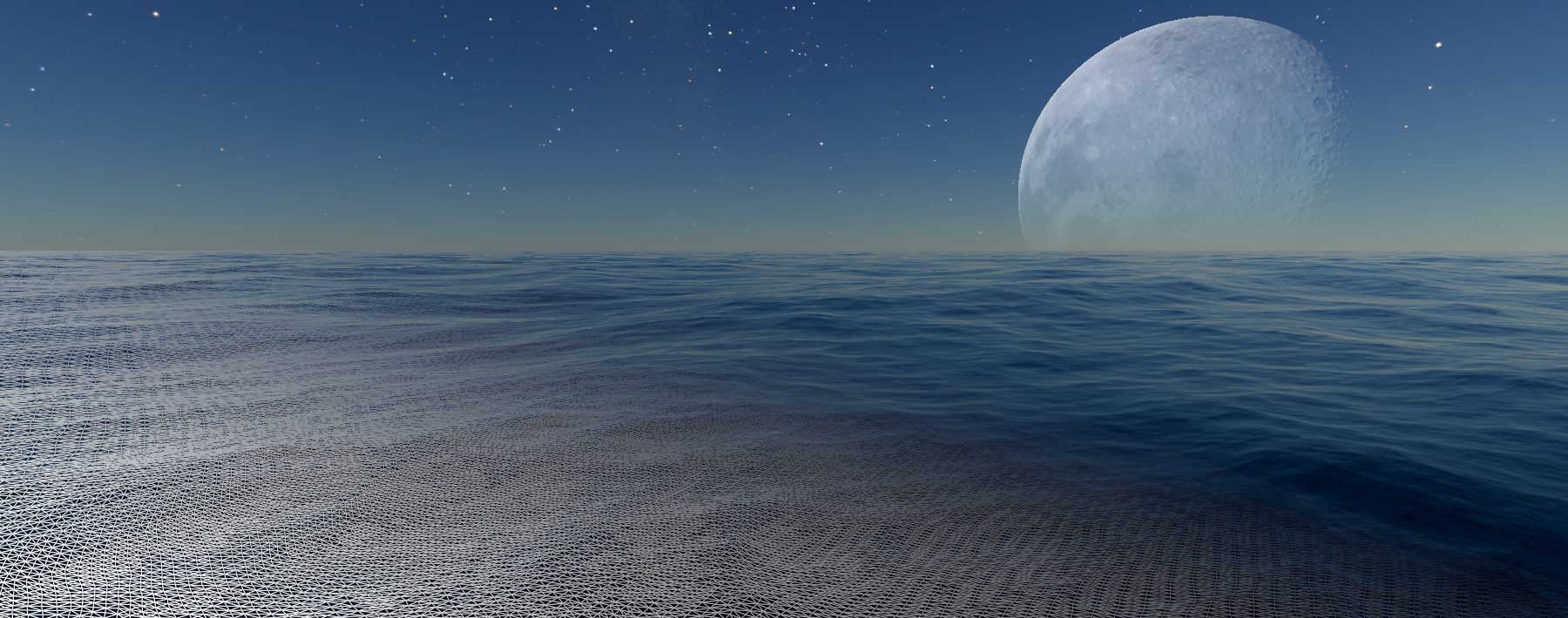}};
    \node[anchor = north west] at (0.0, 0.0) {\includegraphics[fbox, height=0.22\textwidth, trim = 200 0 250 0, clip]{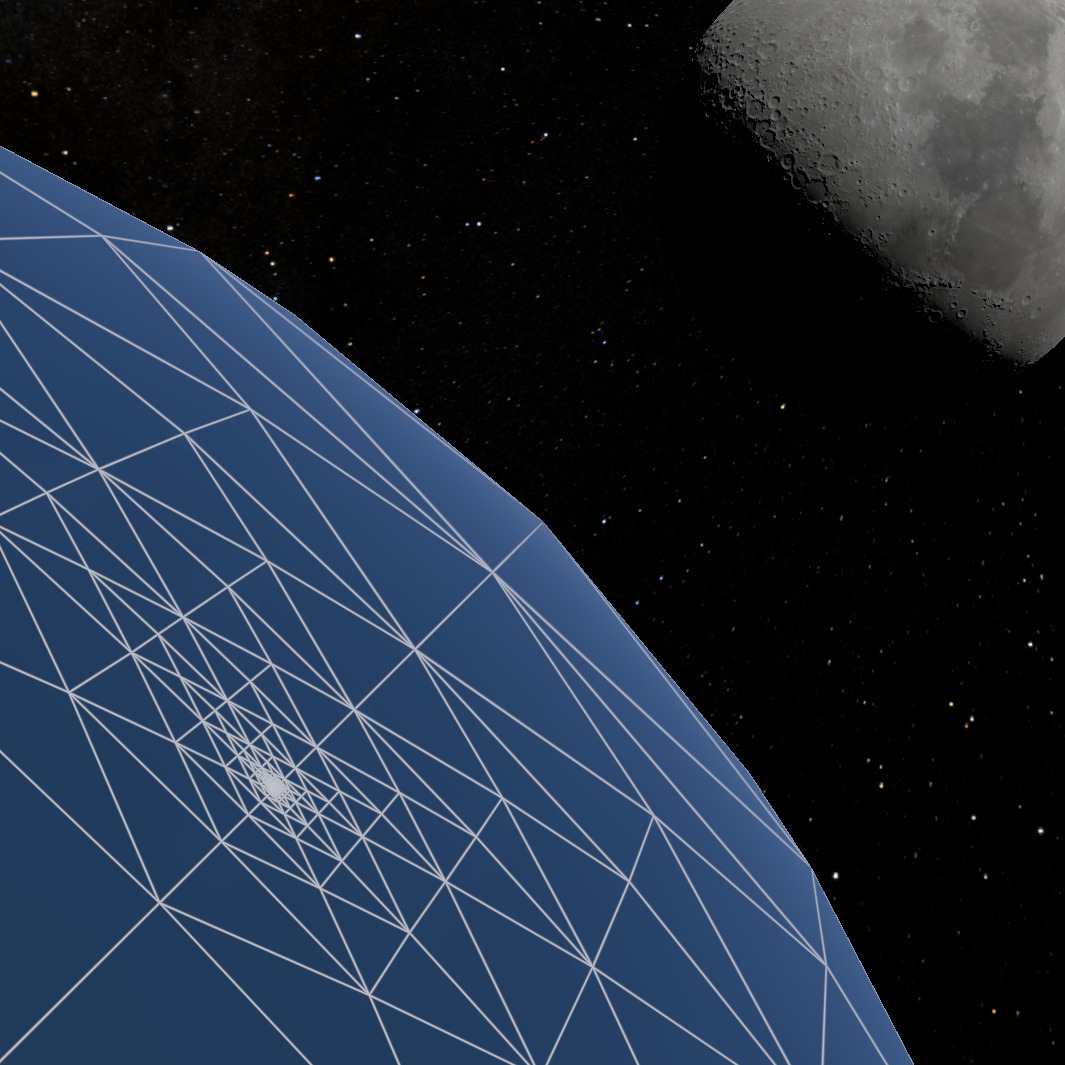}};
    
    \begin{scope}[shift={(-0.18, 0.0)}]
    \node[anchor = north west] at (2.1, 0.0) {\includegraphics[fbox, height=0.22\textwidth, trim = 100 0 480 0, clip]{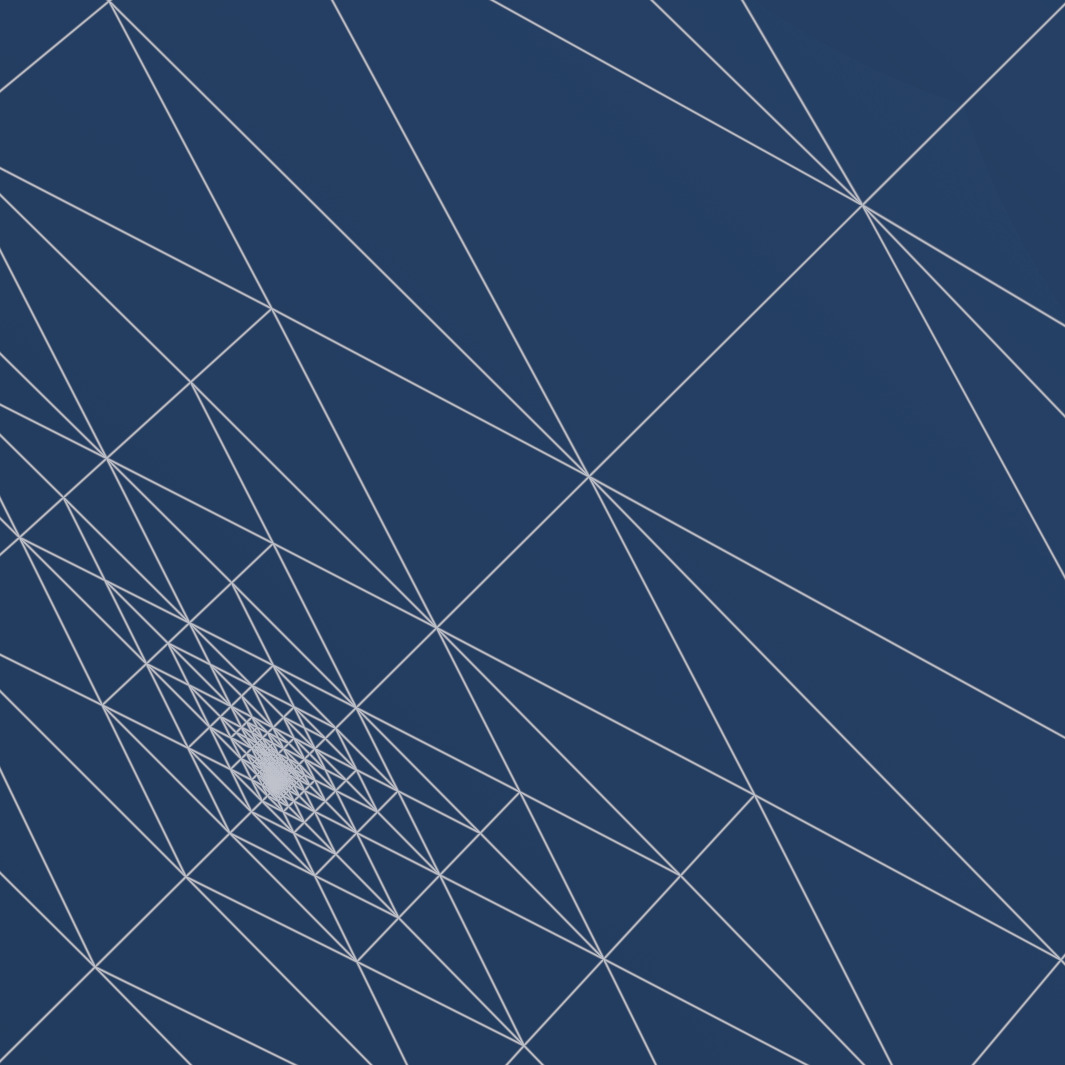}};
    \node[anchor = north west] at (3.65, 0.0) {\includegraphics[fbox, height=0.22\textwidth, trim = 100 0 480 0, clip]{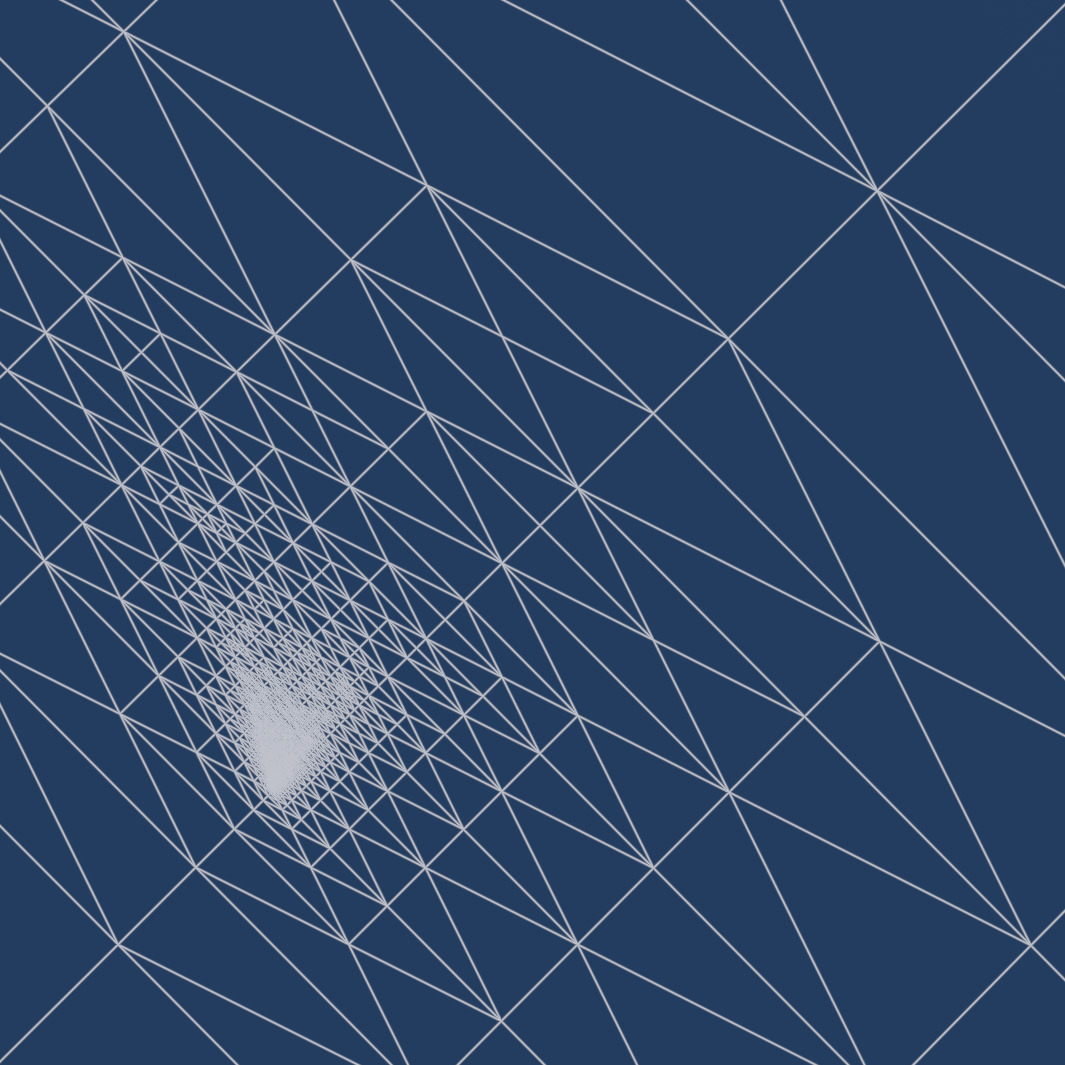}};
    \node[anchor = north west] at (5.2, 0.0) {\includegraphics[fbox, height=0.22\textwidth, trim = 100 0 480 0, clip]{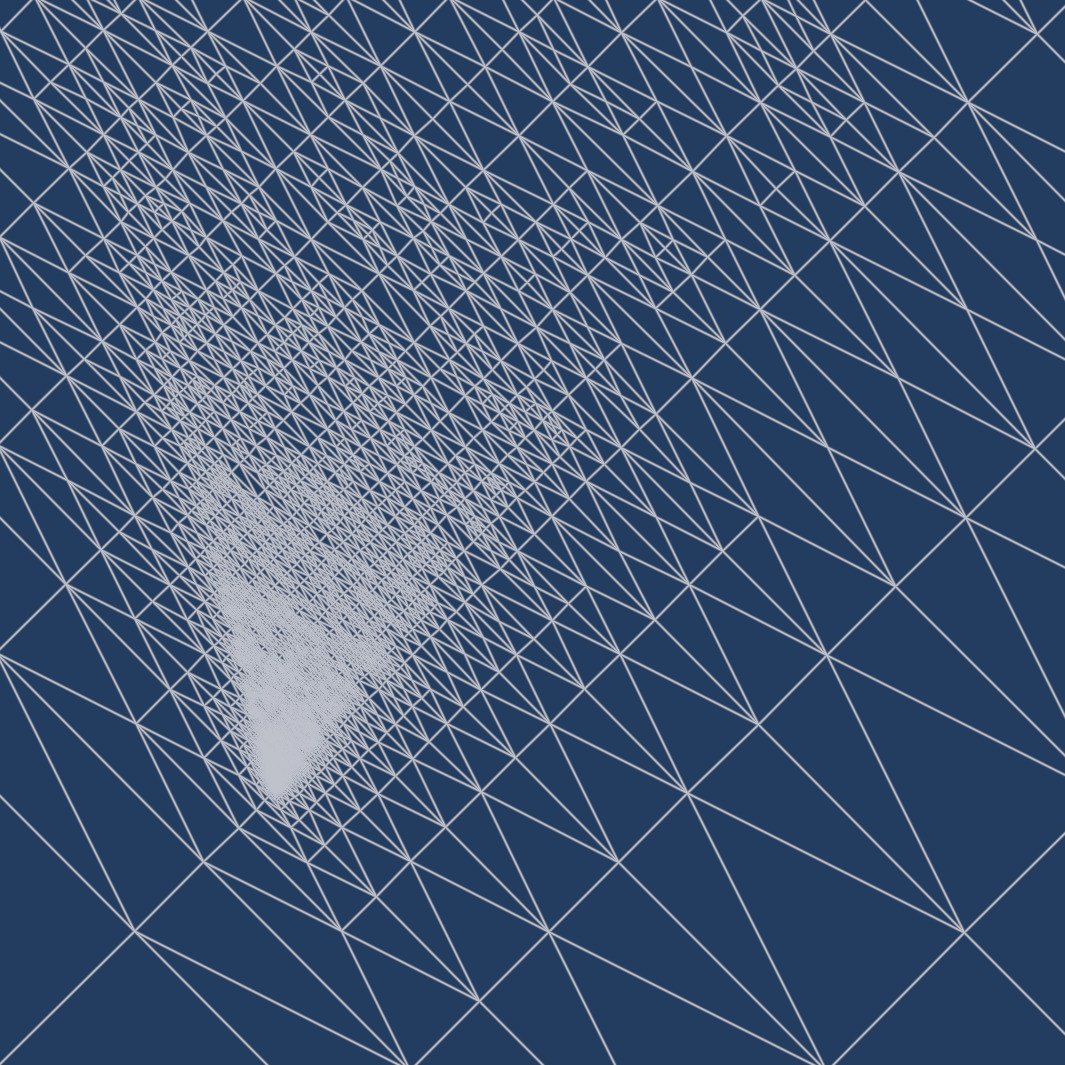}};
    \node[anchor = north west] at (6.75, 0.0) {\includegraphics[fbox, height=0.22\textwidth, trim = 100 0 480 0, clip]{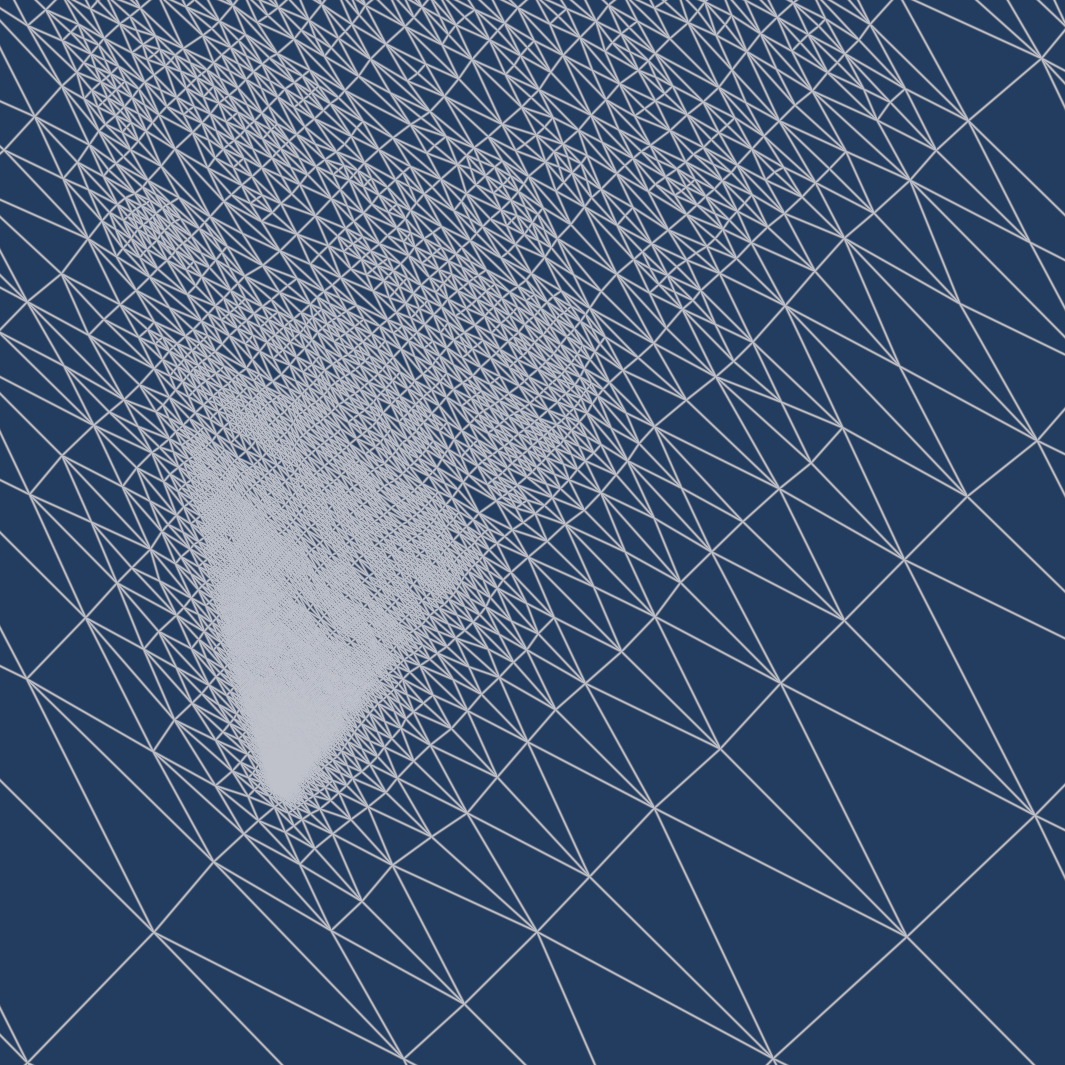}};
    \node[anchor = north west] at (8.3, 0.0) {\includegraphics[fbox, height=0.22\textwidth, trim = 100 0 480 0, clip]{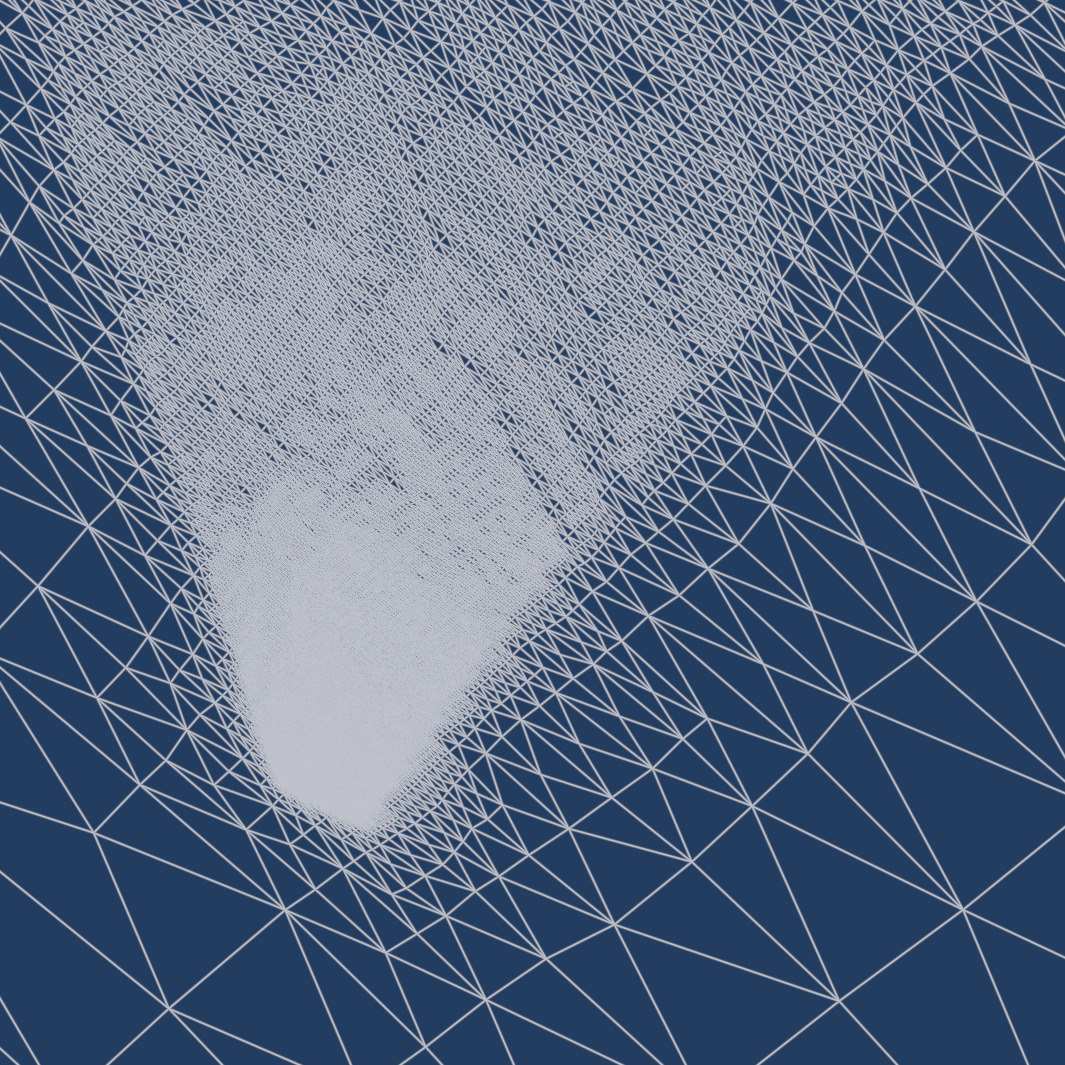}};
    \node[anchor = north west] at (9.85, 0.0) {\includegraphics[fbox, height=0.22\textwidth, trim = 100 0 480 0, clip]{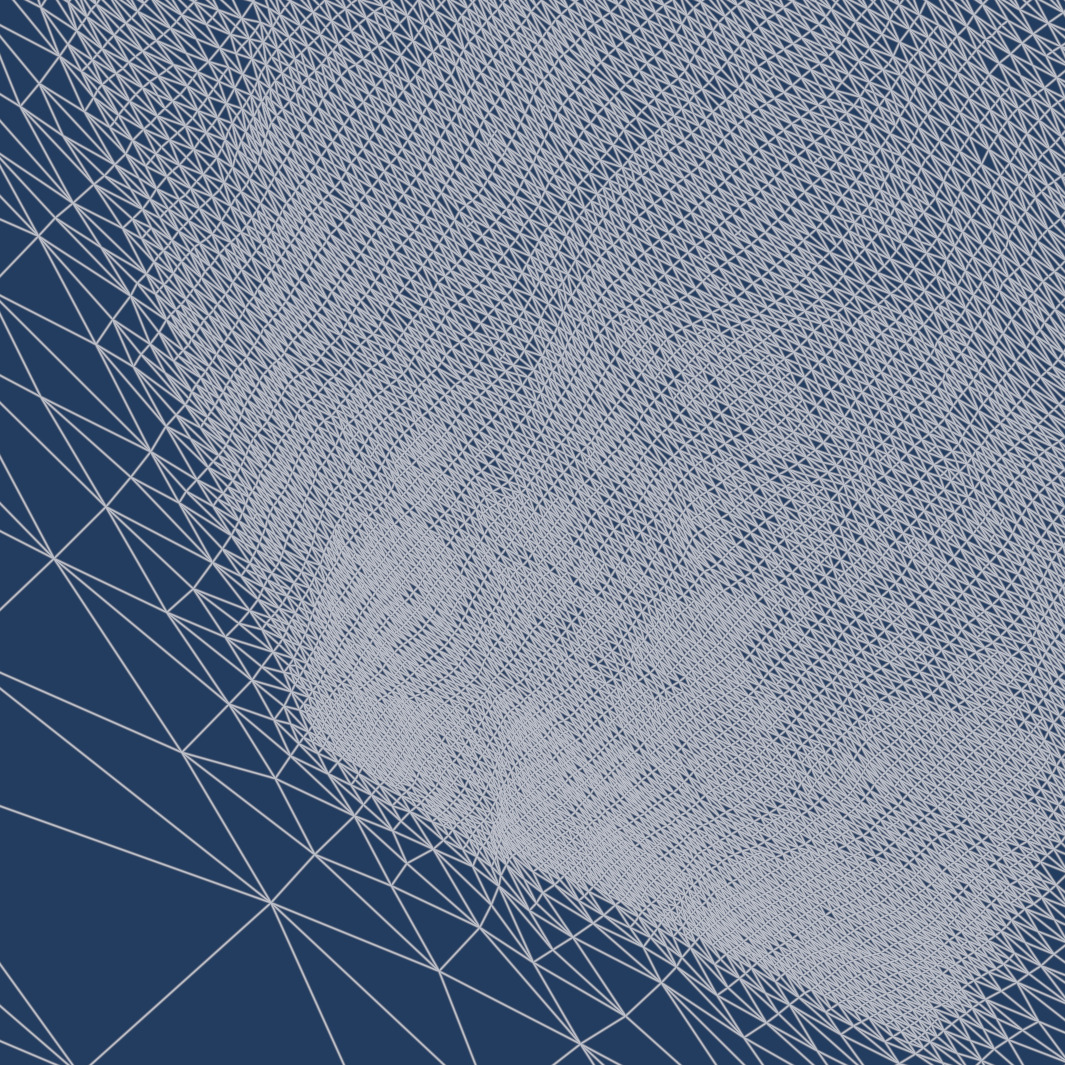}};
    \node[anchor = north east] at (13.0, 0.0) {\includegraphics[fbox, height=0.22\textwidth, trim = 100 0 400 0, clip]{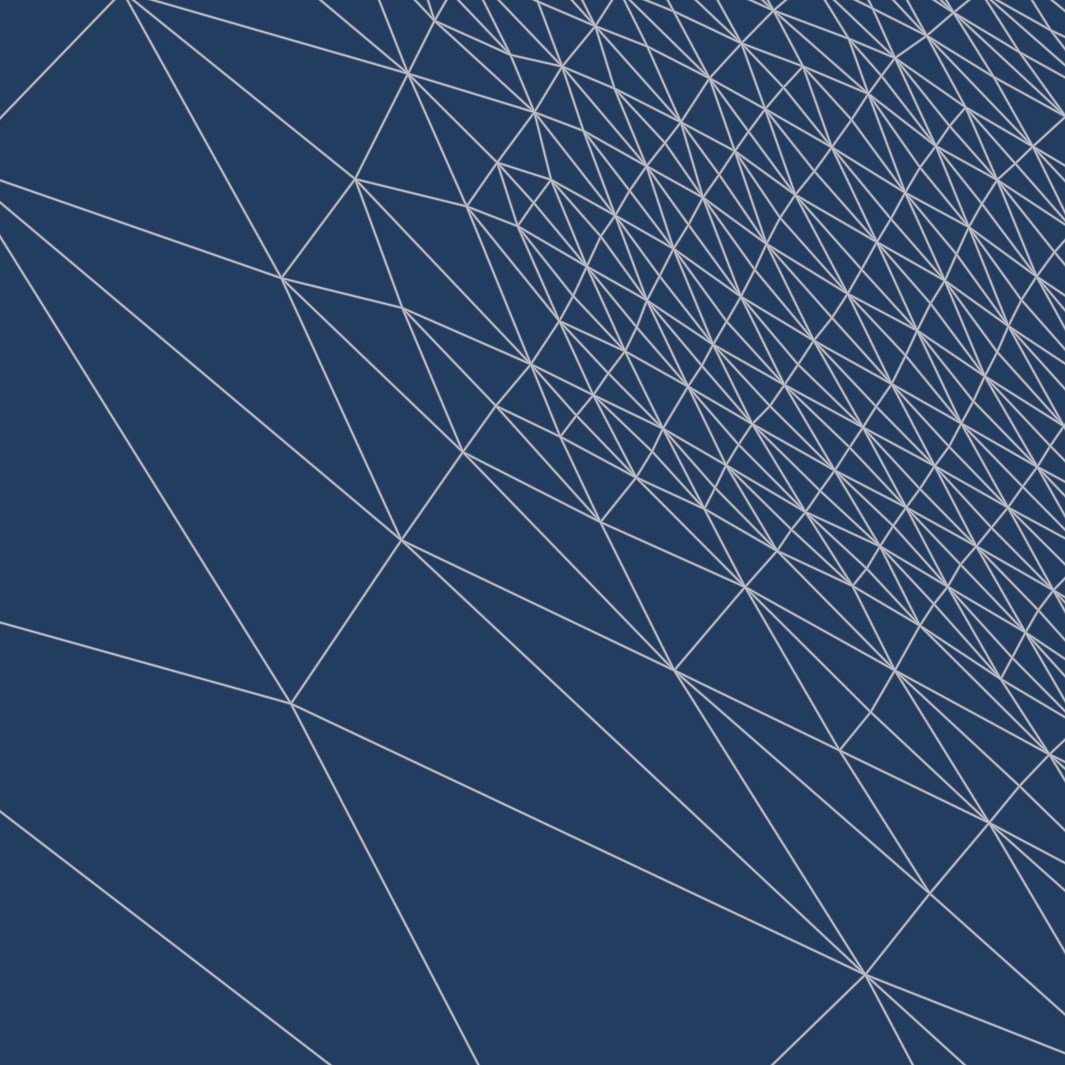}};
    \end{scope}

    \begin{scope}[shift={(0.0, +0.04)}]
        \draw [myred, line width = 1.0]
        (0.122,-2.44) rectangle ++(0.228, 0.46);
        \draw [myred, line width = 1.0, >=stealth, -{Latex[length = 8.000000, width = 5.000000]}]
        (0.35,-2.2) -- (1.4,-2.2)-- (1.4,-1.5)-- (1.92,-1.5);
        \draw [myred, line width = 1.0] 
        (1.92, -3.1) rectangle ++(1.41, 3.06);
    \end{scope}
    
    \begin{scope}[shift={(1.55, +0.04)}]
        \draw [cyan, line width = 1.0] 
        (0.77,-2.43) rectangle ++(0.228, 0.46);

        \draw [cyan, line width = 1.0, >=stealth, -{Latex[length = 8.000000, width = 5.000000]}]
        (0.998,-2.2) -- (1.4,-2.2)-- (1.4,-1.5)-- (1.92,-1.5);

        \draw [cyan, line width = 1.0] 
        (1.92, -3.1) rectangle ++(1.41, 3.06);
    \end{scope}

    \begin{scope}[shift={(3.1, +0.04)}]
        \draw [myred, line width = 1.0] 
        (0.77,-2.43) rectangle ++(0.228, 0.46);
        
        \draw [myred, line width = 1.0, >=stealth, -{Latex[length = 8.000000, width = 5.000000]}]
        (0.998,-2.2) -- (1.4,-2.2)-- (1.4,-1.5)-- (1.92,-1.5);

        \draw [myred, line width = 1.0] 
        (1.92, -3.1) rectangle ++(1.41, 3.06);
    \end{scope}

    \begin{scope}[shift={(4.65, +0.04)}]
        \draw [cyan, line width = 1.0] 
        (0.77,-2.43) rectangle ++(0.228, 0.46);
        
        \draw [cyan, line width = 1.0, >=stealth, -{Latex[length = 8.000000, width = 5.000000]}]
        (0.998,-2.2) -- (1.4,-2.2)-- (1.4,-1.5)-- (1.92,-1.5);

        \draw [cyan, line width = 1.0] 
        (1.92, -3.1) rectangle ++(1.41, 3.06);
    \end{scope}

    \begin{scope}[shift={(6.2, +0.04)}]
        \draw [myred, line width = 1.0] 
        (0.79,-2.43) rectangle ++(0.228, 0.46);
        
        \draw [myred, line width = 1.0, >=stealth, -{Latex[length = 8.000000, width = 5.000000]}]
        (0.79+0.228,-2.2) -- (1.4,-2.2)-- (1.4,-1.5)-- (1.92,-1.5);

        \draw [myred, line width = 1.0] 
        (1.92, -3.1) rectangle ++(1.41, 3.06);
    \end{scope}

    \begin{scope}[shift={(7.75, +0.04)}]
        \draw [cyan, line width = 1.0] 
        (0.77,-2.43) rectangle ++(0.228, 0.46);
        
        \draw [cyan, line width = 1.0, >=stealth, -{Latex[length = 8.000000, width = 5.000000]}]
        (0.998,-2.2) -- (1.4,-2.2)-- (1.4,-1.5)-- (1.92,-1.5);

        \draw [cyan, line width = 1.0] 
        (1.92, -3.1) rectangle ++(1.41, 3.06);
    \end{scope}

    \begin{scope}[shift={(9.3, +0.04)}]
        \draw [myred, line width = 1.0] 
        (0.79,-2.43) rectangle ++(0.228, 0.46);
        
        \draw [myred, line width = 1.0, >=stealth, -{Latex[length = 8.000000, width = 5.000000]}]
        (0.79+0.228,-2.2) -- (1.4,-2.2)-- (1.4,-1.5)-- (1.92,-1.5);

        \draw [myred, line width = 1.0] 
        (1.91, -3.1) rectangle ++(1.6, 3.06);
    \end{scope}
\end{tikzpicture}

%% file: figures/figure-halfedge.tex
\begin{tikzpicture}
    \begin{scope}[scale = 1.0]

    \begin{scope}[shift={(0, 0)}]
        \begin{scope}[scale=3.0]
            \input{figures/figure-halfedge-mesh.tex}
        \end{scope}

        \begin{scope}[shift={(6.6, 0.5)}]
            \fill[mybluelighter, rounded corners] (0.975, -1.6) rectangle ++(0.45, 2.72);
            \fill[mygreenlighter, rounded corners] (-0.55, -1.62) rectangle ++(0.45, 2.72);
            \draw[myblue, line width = 1.0, rounded corners] (0.975, -1.62) rectangle ++(0.45, 2.72);
            \draw[mygreen, line width = 1.0, rounded corners] (-0.55, -1.62) rectangle ++(0.45, 2.72);
            \node (tbl) {
                {\scalebox{0.9}{
                \begin{tabular}{c|cccccccccccc}
                    label & \multicolumn{12}{l}{content} \\
                    \hline
                    \hline
                    $h$ & 0 & 1 & 2 & 3 & 4 & 5 & 6 & 7 & 8 & 9 & 10 & 11 \\
                    \hline
                    \TwinID & 6 & 7 & -1 & -1 & -1 & 8 & 0 & 1 & 5 & -1 & -1 & -1 \\
                \NextID & 1 & 2 & 3 & 0 & 5 & 6 & 4 & 8 & 9 & 10 & 11 & 7 \\
                \PrevID & 3 & 0 & 1 & 2 & 6 & 4 & 5 & 11 & 7 & 8 & 9 & 10\\
                \VertID & 3 & 2 & 1 & 0 & 3 & 4 & 2 & 1 & 2 & 4 & 5 & 6 \\
                \EdgeID & 3 & 4 & 0 & 1 & 2 & 5 & 3 & 4 & 5 & 6 & 7 & 8 \\
                \FaceID & 0 & 0 & 0 & 0 & 1 & 1 & 1 & 2 & 2 & 2 & 2 & 2 \\
            \end{tabular}}}
            };
            
        \end{scope}
    \end{scope}

\end{scope}
\end{tikzpicture}

%% file: figures/figure-bisection-overview.tex
\begin{tikzpicture}
    \begin{scope}[shift={(0.0, 0.0)}]
        \begin{scope}[shift={(+0.0, -0.35)}, scale=3.0]
            \node[mygrey, draw=none, anchor=center] 
            at (0.0, 1.0) {\scalebox{0.7}{\textbf{halfedge mesh (input)}}};
                \input{figures/figure-bisection-overview-mesh.tex}
            \node[black, draw=none, anchor=center] at (0.0, -0.6) {\scalebox{0.75}{(a)}};
        \end{scope}

        \begin{scope}[shift={(4.65, -0.35)}, scale=3.0]
            \node[mygrey, draw=none, anchor=center] 
            at (0.0, 1.0) {\scalebox{0.7}{\textbf{root bisectors}}};
            \input{figures/figure-bisection-overview-root-bisectors.tex}
            \node[black, draw=none, anchor=center] at (0.0, -0.6) {\scalebox{0.75}{(b)}};
        \end{scope}
        
        \begin{scope}[shift={(9.3, -0.35)}, scale=3.0]
            \node[mygrey, draw=none, anchor=center] 
            at (0.0, 1.0) {\scalebox{0.7}{\textbf{adaptive bisection}}};
            \input{figures/figure-bisection-overview-adaptive-bisection.tex}
            \node[black, draw=none, anchor=center] at (0.0, -0.6) {\scalebox{0.75}{(c)}};
        \end{scope}

    \end{scope}
\end{tikzpicture}

%% file: figures/figure-compatibility-chain.tex
\begin{tikzpicture}
    \begin{scope}[shift={(0.0, 0.0)}]
        \begin{scope}[shift={(+0.0, -0.35)}, scale=3.0]
                \input{figures/figure-compatibility-chain-1.tex}
            \node[black, draw=none, anchor=center] at (0.0, -0.6) {\scalebox{0.75}{(a)}};
        \end{scope}

        \begin{scope}[shift={(4.65, -0.35)}, scale=3.0]
            \input{figures/figure-compatibility-chain-2.tex}
            \node[black, draw=none, anchor=center] at (0.0, -0.6) {\scalebox{0.75}{(b)}};
        \end{scope}
        
        \begin{scope}[shift={(9.4, -0.35)}, scale=3.0]
            \input{figures/figure-compatibility-chain-3.tex}
            \node[black, draw=none, anchor=center] at (0.0, -0.6) {\scalebox{0.75}{(c)}};
        \end{scope}
    \end{scope}
\end{tikzpicture}

%% file: figures/figure-cbt.tex
\begin{tikzpicture}
    \begin{scope}[scale = 3.0]
        
    
        \begin{scope}[shift={(0.0,0.0)}, scale=0.14]
            \draw[mygreylight, densely dashed, rounded corners = 1.0, fill = mygreylighterr, anchor = center, line width = 1.0]
            (-0.65, 1.5) rectangle (14.65, 8.50000);

            \draw[black, line width = 0.5]
            (0, 6) -- (0, 7) -- (8, 7) -- (8, 6);
            \draw[black, line width = 0.5]
            (0, 7) -- (0, 8);

            \draw[black, fill = mygreylighterr, line width = 0.5] (0,7) circle (0.25);
            \node[black, draw=none] at (0, 7) {\scalebox{0.7}{$+$}};

            \draw[black, rounded corners = 1.0, fill = mygreylighter, anchor = center, line width = 0.5]
            (-0.500000, 7.650000) rectangle (0.500000, 8.350000);
            \node [black, draw=none] at (0, 8) {\scalebox{0.7}{7}};
            
            \draw[black, line width = 0.5]
            (0, 4) -- (0, 5) -- (4, 5) -- (4, 4);
            \draw[black, line width = 0.5]
            (0, 5) -- (0, 6);
            \draw[black, line width = 0.5]
            (8, 4) -- (8, 5) -- (12, 5) -- (12, 4);
            \draw[black, line width = 0.5]
            (8, 5) -- (8, 6);
            
            \draw[black, fill = mygreylighterr, line width = 0.5] (0,5) circle (0.25);
            \node[black, draw=none] at (0,5) {\scalebox{0.7}{$+$}};
            \draw[black, fill = mygreylighterr, line width = 0.5] (8,5) circle (0.25);
            \node[black, draw=none] at (8, 5) {\scalebox{0.7}{$+$}};
            
            \draw[black, rounded corners = 1.0, fill = mygreylighter, anchor = center, line width = 0.5]
            (-0.500000, 5.650000) rectangle (0.500000, 6.350000);
            \node [black, draw=none] at (0, 6) {\scalebox{0.7}{2}};
            \draw[black, rounded corners = 1.0, fill = mygreylighter, anchor = center, line width = 0.5]
            (7.500000, 5.650000) rectangle (8.500000, 6.350000);
            \node [black, draw=none] at (8, 6) {\scalebox{0.7}{5}};
            \draw[black, line width = 0.5]
            (0, 2) -- (0, 3) -- (2, 3) -- (2, 2);
            \draw[black, line width = 0.5]
            (0, 3) -- (0, 4);
            \draw[black, line width = 0.5]
            (4, 2) -- (4, 3) -- (6, 3) -- (6, 2);
            \draw[black, line width = 0.5]
            (4, 3) -- (4, 4);
            \draw[black, line width = 0.5]
            (8, 2) -- (8, 3) -- (10, 3) -- (10, 2);
            \draw[black, line width = 0.5]
            (8, 3) -- (8, 4);
            \draw[black, line width = 0.5]
            (12, 2) -- (12, 3) -- (14, 3) -- (14, 2);
            \draw[black, line width = 0.5]
            (12, 3) -- (12, 4);

            \draw[black, fill = mygreylighterr, line width = 0.5] (0,3) circle (0.25);
            \node[black, draw=none] at (0, 3) {\scalebox{0.7}{$+$}};
            \draw[black, fill = mygreylighterr, line width = 0.5] (4,3) circle (0.25);
            \node[black, draw=none] at (4,3) {\scalebox{0.7}{$+$}};
            \draw[black, fill = mygreylighterr, line width = 0.5] (8,3) circle (0.25);
            \node[black, draw=none] at (8,3) {\scalebox{0.7}{$+$}};
            \draw[black, fill = mygreylighterr, line width = 0.5] (12,3) circle (0.25);
            \node[black, draw=none] at (12,3) {\scalebox{0.7}{$+$}};

            \draw[black, rounded corners = 1.0, fill = mygreylighter, anchor = center, line width = 0.5]
            (-0.500000, 3.650000) rectangle (0.500000, 4.350000);
            \node [black, draw=none] at (0, 4) {\scalebox{0.7}{2}};
            \draw[black, rounded corners = 1.0, fill = mygreylighter, anchor = center, line width = 0.5]
            (3.500000, 3.650000) rectangle (4.500000, 4.350000);
            \node [black, draw=none] at (4, 4) {\scalebox{0.7}{0}};
            \draw[black, rounded corners = 1.0, fill = mygreylighter, anchor = center, line width = 0.5]
            (7.500000, 3.650000) rectangle (8.500000, 4.350000);
            \node [black, draw=none] at (8, 4) {\scalebox{0.7}{2}};
            \draw[black, rounded corners = 1.0, fill = mygreylighter, anchor = center, line width = 0.5]
            (11.500000, 3.650000) rectangle (12.500000, 4.350000);
            \node [black, draw=none] at (12, 4) {\scalebox{0.7}{3}};
            \draw[black, line width = 0.5]
            (0, 0) -- (0, 1) -- (1, 1) -- (1, 0);
            \draw[black, line width = 0.5]
            (0, 1) -- (0, 2);
            \draw[black, line width = 0.5]
            (2, 0) -- (2, 1) -- (3, 1) -- (3, 0);
            \draw[black, line width = 0.5]
            (2, 1) -- (2, 2);
            \draw[black, line width = 0.5]
            (4, 0) -- (4, 1) -- (5, 1) -- (5, 0);
            \draw[black, line width = 0.5]
            (4, 1) -- (4, 2);
            \draw[black, line width = 0.5]
            (6, 0) -- (6, 1) -- (7, 1) -- (7, 0);
            \draw[black, line width = 0.5]
            (6, 1) -- (6, 2);
            \draw[black, line width = 0.5]
            (8, 0) -- (8, 1) -- (9, 1) -- (9, 0);
            \draw[black, line width = 0.5]
            (8, 1) -- (8, 2);
            \draw[black, line width = 0.5]
            (10, 0) -- (10, 1) -- (11, 1) -- (11, 0);
            \draw[black, line width = 0.5]
            (10, 1) -- (10, 2);
            \draw[black, line width = 0.5]
            (12, 0) -- (12, 1) -- (13, 1) -- (13, 0);
            \draw[black, line width = 0.5]
            (12, 1) -- (12, 2);
            \draw[black, line width = 0.5]
            (14, 0) -- (14, 1) -- (15, 1) -- (15, 0);
            \draw[black, line width = 0.5]
            (14, 1) -- (14, 2);

            \draw[black, fill = mygreylighterr, line width = 0.5] (0,1) circle (0.25);
            \node[black, draw=none] at (0,1) {\scalebox{0.7}{$+$}};
            \draw[black, fill = mygreylighterr, line width = 0.5] (2,1) circle (0.25);
            \node[black, draw=none] at (2,1) {\scalebox{0.7}{$+$}};
            \draw[black, fill = mygreylighterr, line width = 0.5] (4,1) circle (0.25);
            \node[black, draw=none] at (4,1) {\scalebox{0.7}{$+$}};
            \draw[black, fill = mygreylighterr, line width = 0.5] (6,1) circle (0.25);
            \node[black, draw=none] at (6,1) {\scalebox{0.7}{$+$}};
            \draw[black, fill = mygreylighterr, line width = 0.5] (8,1) circle (0.25);
            \node[black, draw=none] at (8,1) {\scalebox{0.7}{$+$}};
            \draw[black, fill = mygreylighterr, line width = 0.5] (10,1) circle (0.25);
            \node[black, draw=none] at (10,1) {\scalebox{0.7}{$+$}};
            \draw[black, fill = mygreylighterr, line width = 0.5] (12,1) circle (0.25);
            \node[black, draw=none] at (12,1) {\scalebox{0.7}{$+$}};
            \draw[black, fill = mygreylighterr, line width = 0.5] (14,1) circle (0.25);
            \node[black, draw=none] at (14,1) {\scalebox{0.7}{$+$}};

            \draw[black, rounded corners = 1.0, fill = mygreylighter, anchor = center, line width = 0.5]
            (-0.500000, 1.650000) rectangle (0.500000, 2.350000);
            \node [black, draw=none] at (0, 2) {\scalebox{0.7}{1}};
            \draw[black, rounded corners = 1.0, fill = mygreylighter, anchor = center, line width = 0.5]
            (1.500000, 1.650000) rectangle (2.500000, 2.350000);
            \node [black, draw=none] at (2, 2) {\scalebox{0.7}{1}};
            \draw[black, rounded corners = 1.0, fill = mygreylighter, anchor = center, line width = 0.5]
            (3.500000, 1.650000) rectangle (4.500000, 2.350000);
            \node [black, draw=none] at (4, 2) {\scalebox{0.7}{0}};
            \draw[black, rounded corners = 1.0, fill = mygreylighter, anchor = center, line width = 0.5]
            (5.500000, 1.650000) rectangle (6.500000, 2.350000);
            \node [black, draw=none] at (6, 2) {\scalebox{0.7}{0}};
            \draw[black, rounded corners = 1.0, fill = mygreylighter, anchor = center, line width = 0.5]
            (7.500000, 1.650000) rectangle (8.500000, 2.350000);
            \node [black, draw=none] at (8, 2) {\scalebox{0.7}{0}};
            \draw[black, rounded corners = 1.0, fill = mygreylighter, anchor = center, line width = 0.5]
            (9.500000, 1.650000) rectangle (10.500000, 2.350000);
            \node [black, draw=none] at (10, 2) {\scalebox{0.7}{2}};
            \draw[black, rounded corners = 1.0, fill = mygreylighter, anchor = center, line width = 0.5]
            (11.500000, 1.650000) rectangle (12.500000, 2.350000);
            \node [black, draw=none] at (12, 2) {\scalebox{0.7}{2}};
            \draw[black, rounded corners = 1.0, fill = mygreylighter, anchor = center, line width = 0.5]
            (13.500000, 1.650000) rectangle (14.500000, 2.350000);
            \node [black, draw=none] at (14, 2) {\scalebox{0.7}{1}};
            \draw[black, rounded corners = 1.0, fill = mygreenlight, anchor = center, line width = 0.5]
            (-0.500000, -0.350000) rectangle (0.500000, 0.350000);
            \node [black, draw=none] at (0, 0) {\scalebox{0.7}{1}};
            \draw[black, rounded corners = 1.0, fill = mygreenlight, anchor = center, line width = 0.5]
            (0.500000, -0.350000) rectangle (1.500000, 0.350000);
            \node [black, draw=none] at (1, 0) {\scalebox{0.7}{0}};
            \draw[black, rounded corners = 1.0, fill = mygreenlight, anchor = center, line width = 0.5]
            (1.500000, -0.350000) rectangle (2.500000, 0.350000);
            \node [black, draw=none] at (2, 0) {\scalebox{0.7}{0}};
            \draw[black, rounded corners = 1.0, fill = mygreenlight, anchor = center, line width = 0.5]
            (2.500000, -0.350000) rectangle (3.500000, 0.350000);
            \node [black, draw=none] at (3, 0) {\scalebox{0.7}{1}};
            \draw[black, rounded corners = 1.0, fill = mygreenlight, anchor = center, line width = 0.5]
            (3.500000, -0.350000) rectangle (4.500000, 0.350000);
            \node [black, draw=none] at (4, 0) {\scalebox{0.7}{0}};
            \draw[black, rounded corners = 1.0, fill = mygreenlight, anchor = center, line width = 0.5]
            (4.500000, -0.350000) rectangle (5.500000, 0.350000);
            \node [black, draw=none] at (5, 0) {\scalebox{0.7}{0}};
            \draw[black, rounded corners = 1.0, fill = mygreenlight, anchor = center, line width = 0.5]
            (5.500000, -0.350000) rectangle (6.500000, 0.350000);
            \node [black, draw=none] at (6, 0) {\scalebox{0.7}{0}};
            \draw[black, rounded corners = 1.0, fill = mygreenlight, anchor = center, line width = 0.5]
            (6.500000, -0.350000) rectangle (7.500000, 0.350000);
            \node [black, draw=none] at (7, 0) {\scalebox{0.7}{0}};
            \draw[black, rounded corners = 1.0, fill = mygreenlight, anchor = center, line width = 0.5]
            (7.500000, -0.350000) rectangle (8.500000, 0.350000);
            \node [black, draw=none] at (8, 0) {\scalebox{0.7}{0}};
            \draw[black, rounded corners = 1.0, fill = mygreenlight, anchor = center, line width = 0.5]
            (8.500000, -0.350000) rectangle (9.500000, 0.350000);
            \node [black, draw=none] at (9, 0) {\scalebox{0.7}{0}};
            \draw[black, rounded corners = 1.0, fill = mygreenlight, anchor = center, line width = 0.5]
            (9.500000, -0.350000) rectangle (10.500000, 0.350000);
            \node [black, draw=none] at (10, 0) {\scalebox{0.7}{1}};
            \draw[black, rounded corners = 1.0, fill = mygreenlight, anchor = center, line width = 0.5]
            (10.500000, -0.350000) rectangle (11.500000, 0.350000);
            \node [black, draw=none] at (11, 0) {\scalebox{0.7}{1}};
            \draw[black, rounded corners = 1.0, fill = mygreenlight, anchor = center, line width = 0.5]
            (11.500000, -0.350000) rectangle (12.500000, 0.350000);
            \node [black, draw=none] at (12, 0) {\scalebox{0.7}{1}};
            \draw[black, rounded corners = 1.0, fill = mygreenlight, anchor = center, line width = 0.5]
            (12.500000, -0.350000) rectangle (13.500000, 0.350000);
            \node [black, draw=none] at (13, 0) {\scalebox{0.7}{1}};
            \draw[black, rounded corners = 1.0, fill = mygreenlight, anchor = center, line width = 0.5]
            (13.500000, -0.350000) rectangle (14.500000, 0.350000);
            \node [black, draw=none] at (14, 0) {\scalebox{0.7}{1}};
            \draw[black, rounded corners = 1.0, fill = mygreenlight, anchor = center, line width = 0.5]
            (14.500000, -0.350000) rectangle (15.500000, 0.350000);
            \node [black, draw=none] at (15, 0) {\scalebox{0.7}{0}};
    
            \node [anchor = west, black, draw=none] at (-0.75, +9.00) {\scalebox{0.75}{CBT data:}};
            \node [anchor = west, mygreen, draw=none] at (-0.75, -0.75) {\scalebox{0.75}{bitfield}};
            \node [anchor = west, mygrey, draw=none] at (9.0, 8.0) {\scalebox{0.75}{sum-reduction tree}};
        \end{scope}
    
        \begin{scope}[shift={(2.4,0.0)}, scale=0.14]
            \draw[mygreylight, line width = 0.5]
            (0, 6) -- (0, 7) -- (8, 7) -- (8, 6);
            \draw[mygreylight, line width = 0.5]
            (0, 7) -- (0, 8);

            \draw[mygreylight, fill = mygreylighterr, line width = 0.5] (0,7) circle (0.25);
            \node[mygreylight, draw=none] at (0, 7) {\scalebox{0.7}{$+$}};

            \draw[black, rounded corners = 1.0, fill = mygreylighter, anchor = center, line width = 0.5]
            (-0.500000, 7.650000) rectangle (0.500000, 8.350000);
            \node [black, draw=none] at (0, 8) {\scalebox{0.7}{1}};
            \draw[mygreylight, line width = 0.5]
            (0, 4) -- (0, 5) -- (4, 5) -- (4, 4);
            \draw[mygreylight, line width = 0.5]
            (0, 5) -- (0, 6);
            \draw[mygreylight, line width = 0.5]
            (8, 4) -- (8, 5) -- (12, 5) -- (12, 4);
            \draw[mygreylight, line width = 0.5]
            (8, 5) -- (8, 6);

            \draw[mygreylight, fill = mygreylighterr, line width = 0.5] (0,5) circle (0.25);
            \node[mygreylight, draw=none] at (0,5) {\scalebox{0.7}{$+$}};
            \draw[mygreylight, fill = mygreylighterr, line width = 0.5] (8,5) circle (0.25);
            \node[mygreylight, draw=none] at (8, 5) {\scalebox{0.7}{$+$}};

            \draw[black, rounded corners = 1.0, fill = mygreylighter, anchor = center, line width = 0.5]
            (-0.500000, 5.650000) rectangle (0.500000, 6.350000);
            \node [black, draw=none] at (0, 6) {\scalebox{0.7}{2}};
            \draw[black, rounded corners = 1.0, fill = mygreylighter, anchor = center, line width = 0.5]
            (7.500000, 5.650000) rectangle (8.500000, 6.350000);
            \node [black, draw=none] at (8, 6) {\scalebox{0.7}{3}};
            \draw[mygreylight, line width = 0.5]
            (0, 2) -- (0, 3) -- (2, 3) -- (2, 2);
            \draw[mygreylight, line width = 0.5]
            (0, 3) -- (0, 4);
            \draw[mygreylight, line width = 0.5]
            (4, 2) -- (4, 3) -- (6, 3) -- (6, 2);
            \draw[mygreylight, line width = 0.5]
            (4, 3) -- (4, 4);
            \draw[mygreylight, line width = 0.5]
            (8, 2) -- (8, 3) -- (10, 3) -- (10, 2);
            \draw[mygreylight, line width = 0.5]
            (8, 3) -- (8, 4);
            \draw[mygreylight, line width = 0.5]
            (12, 2) -- (12, 3) -- (14, 3) -- (14, 2);
            \draw[mygreylight, line width = 0.5]
            (12, 3) -- (12, 4);

            \draw[mygreylight, fill = mygreylighterr, line width = 0.5] (0,3) circle (0.25);
            \node[mygreylight, draw=none] at (0, 3) {\scalebox{0.7}{$+$}};
            \draw[mygreylight, fill = mygreylighterr, line width = 0.5] (4,3) circle (0.25);
            \node[mygreylight, draw=none] at (4,3) {\scalebox{0.7}{$+$}};
            \draw[mygreylight, fill = mygreylighterr, line width = 0.5] (8,3) circle (0.25);
            \node[mygreylight, draw=none] at (8,3) {\scalebox{0.7}{$+$}};
            \draw[mygreylight, fill = mygreylighterr, line width = 0.5] (12,3) circle (0.25);
            \node[mygreylight, draw=none] at (12,3) {\scalebox{0.7}{$+$}};

            \draw[black, rounded corners = 1.0, fill = mygreylighter, anchor = center, line width = 0.5]
            (-0.500000, 3.650000) rectangle (0.500000, 4.350000);
            \node [black, draw=none] at (0, 4) {\scalebox{0.7}{4}};
            \draw[black, rounded corners = 1.0, fill = mygreylighter, anchor = center, line width = 0.5]
            (3.500000, 3.650000) rectangle (4.500000, 4.350000);
            \node [black, draw=none] at (4, 4) {\scalebox{0.7}{5}};
            \draw[black, rounded corners = 1.0, fill = mygreylighter, anchor = center, line width = 0.5]
            (7.500000, 3.650000) rectangle (8.500000, 4.350000);
            \node [black, draw=none] at (8, 4) {\scalebox{0.7}{6}};
            \draw[black, rounded corners = 1.0, fill = mygreylighter, anchor = center, line width = 0.5]
            (11.500000, 3.650000) rectangle (12.500000, 4.350000);
            \node [black, draw=none] at (12, 4) {\scalebox{0.7}{7}};
            \draw[mygreylight, line width = 0.5]
            (0, 0) -- (0, 1) -- (1, 1) -- (1, 0);
            \draw[mygreylight, line width = 0.5]
            (0, 1) -- (0, 2);
            \draw[mygreylight, line width = 0.5]
            (2, 0) -- (2, 1) -- (3, 1) -- (3, 0);
            \draw[mygreylight, line width = 0.5]
            (2, 1) -- (2, 2);
            \draw[mygreylight, line width = 0.5]
            (4, 0) -- (4, 1) -- (5, 1) -- (5, 0);
            \draw[mygreylight, line width = 0.5]
            (4, 1) -- (4, 2);
            \draw[mygreylight, line width = 0.5]
            (6, 0) -- (6, 1) -- (7, 1) -- (7, 0);
            \draw[mygreylight, line width = 0.5]
            (6, 1) -- (6, 2);
            \draw[mygreylight, line width = 0.5]
            (8, 0) -- (8, 1) -- (9, 1) -- (9, 0);
            \draw[mygreylight, line width = 0.5]
            (8, 1) -- (8, 2);
            \draw[mygreylight, line width = 0.5]
            (10, 0) -- (10, 1) -- (11, 1) -- (11, 0);
            \draw[mygreylight, line width = 0.5]
            (10, 1) -- (10, 2);
            \draw[mygreylight, line width = 0.5]
            (12, 0) -- (12, 1) -- (13, 1) -- (13, 0);
            \draw[mygreylight, line width = 0.5]
            (12, 1) -- (12, 2);
            \draw[mygreylight, line width = 0.5]
            (14, 0) -- (14, 1) -- (15, 1) -- (15, 0);
            \draw[mygreylight, line width = 0.5]
            (14, 1) -- (14, 2);

            \draw[mygreylight, fill = mygreylighterr, line width = 0.5] (0,1) circle (0.25);
            \node[mygreylight, draw=none] at (0,1) {\scalebox{0.7}{$+$}};
            \draw[mygreylight, fill = mygreylighterr, line width = 0.5] (2,1) circle (0.25);
            \node[mygreylight, draw=none] at (2,1) {\scalebox{0.7}{$+$}};
            \draw[mygreylight, fill = mygreylighterr, line width = 0.5] (4,1) circle (0.25);
            \node[mygreylight, draw=none] at (4,1) {\scalebox{0.7}{$+$}};
            \draw[mygreylight, fill = mygreylighterr, line width = 0.5] (6,1) circle (0.25);
            \node[mygreylight, draw=none] at (6,1) {\scalebox{0.7}{$+$}};
            \draw[mygreylight, fill = mygreylighterr, line width = 0.5] (8,1) circle (0.25);
            \node[mygreylight, draw=none] at (8,1) {\scalebox{0.7}{$+$}};
            \draw[mygreylight, fill = mygreylighterr, line width = 0.5] (10,1) circle (0.25);
            \node[mygreylight, draw=none] at (10,1) {\scalebox{0.7}{$+$}};
            \draw[mygreylight, fill = mygreylighterr, line width = 0.5] (12,1) circle (0.25);
            \node[mygreylight, draw=none] at (12,1) {\scalebox{0.7}{$+$}};
            \draw[mygreylight, fill = mygreylighterr, line width = 0.5] (14,1) circle (0.25);
            \node[mygreylight, draw=none] at (14,1) {\scalebox{0.7}{$+$}};

            \draw[black, rounded corners = 1.0, fill = mygreylighter, anchor = center, line width = 0.5]
            (-0.500000, 1.650000) rectangle (0.500000, 2.350000);
            \node [black, draw=none] at (0, 2) {\scalebox{0.7}{8}};
            \draw[black, rounded corners = 1.0, fill = mygreylighter, anchor = center, line width = 0.5]
            (1.500000, 1.650000) rectangle (2.500000, 2.350000);
            \node [black, draw=none] at (2, 2) {\scalebox{0.7}{9}};
            \draw[black, rounded corners = 1.0, fill = mygreylighter, anchor = center, line width = 0.5]
            (3.500000, 1.650000) rectangle (4.500000, 2.350000);
            \node [black, draw=none] at (4, 2) {\scalebox{0.7}{10}};
            \draw[black, rounded corners = 1.0, fill = mygreylighter, anchor = center, line width = 0.5]
            (5.500000, 1.650000) rectangle (6.500000, 2.350000);
            \node [black, draw=none] at (6, 2) {\scalebox{0.7}{11}};
            \draw[black, rounded corners = 1.0, fill = mygreylighter, anchor = center, line width = 0.5]
            (7.500000, 1.650000) rectangle (8.500000, 2.350000);
            \node [black, draw=none] at (8, 2) {\scalebox{0.7}{12}};
            \draw[black, rounded corners = 1.0, fill = mygreylighter, anchor = center, line width = 0.5]
            (9.500000, 1.650000) rectangle (10.500000, 2.350000);
            \node [black, draw=none] at (10, 2) {\scalebox{0.7}{13}};
            \draw[black, rounded corners = 1.0, fill = mygreylighter, anchor = center, line width = 0.5]
            (11.500000, 1.650000) rectangle (12.500000, 2.350000);
            \node [black, draw=none] at (12, 2) {\scalebox{0.7}{14}};
            \draw[black, rounded corners = 1.0, fill = mygreylighter, anchor = center, line width = 0.5]
            (13.500000, 1.650000) rectangle (14.500000, 2.350000);
            \node [black, draw=none] at (14, 2) {\scalebox{0.7}{15}};
            \draw[black, rounded corners = 1.0, fill = mygreenlight, anchor = center, line width = 0.5]
            (-0.500000, -0.350000) rectangle (0.500000, 0.350000);
            \node [black, draw=none] at (0, 0) {\scalebox{0.7}{16}};
            \draw[black, rounded corners = 1.0, fill = mygreenlight, anchor = center, line width = 0.5]
            (0.500000, -0.350000) rectangle (1.500000, 0.350000);
            \node [black, draw=none] at (1, 0) {\scalebox{0.7}{17}};
            \draw[black, rounded corners = 1.0, fill = mygreenlight, anchor = center, line width = 0.5]
            (1.500000, -0.350000) rectangle (2.500000, 0.350000);
            \node [black, draw=none] at (2, 0) {\scalebox{0.7}{18}};
            \draw[black, rounded corners = 1.0, fill = mygreenlight, anchor = center, line width = 0.5]
            (2.500000, -0.350000) rectangle (3.500000, 0.350000);
            \node [black, draw=none] at (3, 0) {\scalebox{0.7}{19}};
            \draw[black, rounded corners = 1.0, fill = mygreenlight, anchor = center, line width = 0.5]
            (3.500000, -0.350000) rectangle (4.500000, 0.350000);
            \node [black, draw=none] at (4, 0) {\scalebox{0.7}{20}};
            \draw[black, rounded corners = 1.0, fill = mygreenlight, anchor = center, line width = 0.5]
            (4.500000, -0.350000) rectangle (5.500000, 0.350000);
            \node [black, draw=none] at (5, 0) {\scalebox{0.7}{21}};
            \draw[black, rounded corners = 1.0, fill = mygreenlight, anchor = center, line width = 0.5]
            (5.500000, -0.350000) rectangle (6.500000, 0.350000);
            \node [black, draw=none] at (6, 0) {\scalebox{0.7}{22}};
            \draw[black, rounded corners = 1.0, fill = mygreenlight, anchor = center, line width = 0.5]
            (6.500000, -0.350000) rectangle (7.500000, 0.350000);
            \node [black, draw=none] at (7, 0) {\scalebox{0.7}{23}};
            \draw[black, rounded corners = 1.0, fill = mygreenlight, anchor = center, line width = 0.5]
            (7.500000, -0.350000) rectangle (8.500000, 0.350000);
            \node [black, draw=none] at (8, 0) {\scalebox{0.7}{24}};
            \draw[black, rounded corners = 1.0, fill = mygreenlight, anchor = center, line width = 0.5]
            (8.500000, -0.350000) rectangle (9.500000, 0.350000);
            \node [black, draw=none] at (9, 0) {\scalebox{0.7}{25}};
            \draw[black, rounded corners = 1.0, fill = mygreenlight, anchor = center, line width = 0.5]
            (9.500000, -0.350000) rectangle (10.500000, 0.350000);
            \node [black, draw=none] at (10, 0) {\scalebox{0.7}{26}};
            \draw[black, rounded corners = 1.0, fill = mygreenlight, anchor = center, line width = 0.5]
            (10.500000, -0.350000) rectangle (11.500000, 0.350000);
            \node [black, draw=none] at (11, 0) {\scalebox{0.7}{27}};
            \draw[black, rounded corners = 1.0, fill = mygreenlight, anchor = center, line width = 0.5]
            (11.500000, -0.350000) rectangle (12.500000, 0.350000);
            \node [black, draw=none] at (12, 0) {\scalebox{0.7}{28}};
            \draw[black, rounded corners = 1.0, fill = mygreenlight, anchor = center, line width = 0.5]
            (12.500000, -0.350000) rectangle (13.500000, 0.350000);
            \node [black, draw=none] at (13, 0) {\scalebox{0.7}{29}};
            \draw[black, rounded corners = 1.0, fill = mygreenlight, anchor = center, line width = 0.5]
            (13.500000, -0.350000) rectangle (14.500000, 0.350000);
            \node [black, draw=none] at (14, 0) {\scalebox{0.7}{30}};
            \draw[black, rounded corners = 1.0, fill = mygreenlight, anchor = center, line width = 0.5]
            (14.500000, -0.350000) rectangle (15.500000, 0.350000);
            \node [black, draw=none] at (15, 0) {\scalebox{0.7}{31}};

            \draw[black, line width = 0.5, ->, >=stealth]
            (0.5, 8) -- (0.75, 8) -- (0.75, 7.5) -- (-0.9, 7.5) -- (-0.9, 6.0) -- (-0.5, 6.0);
            \draw[black, line width = 0.5, ->, >=stealth]
            (0.5, 6.0) -- (7.5, 6.0);
            \draw[black, line width = 0.5, ->, >=stealth]
            (8.5, 6) -- (8.75, 6) -- (8.75, 5.5) -- (-0.9, 5.5) -- (-0.9, 4.0) -- (-0.5, 4.0);
            \draw[black, line width = 0.5, ->, >=stealth] (0.5, 4.0) -- (3.5, 4.0);
            \draw[black, line width = 0.5, ->, >=stealth] (4.5, 4.0) -- (7.5, 4.0);
            \draw[black, line width = 0.5, ->, >=stealth] (8.5, 4.0) -- (11.5, 4.0);
            \draw[black, line width = 0.5, ->, >=stealth]
            (12.5, 4) -- (12.75, 4) -- (12.75, 3.5) -- (-0.9, 3.5) -- (-0.9, 2.0) -- (-0.5, 2.0);
            \draw[black, line width = 0.5, ->, >=stealth]
            (14.5, 2) -- (14.75, 2) -- (14.75, 1.5) -- (-0.9, 1.5) -- (-0.9, 0.0) -- (-0.5, 0.0);

            \node [anchor = west, black, draw=none] at (-0.75, +9.00) {\scalebox{0.75}{array layout:}};
        \end{scope}
    
    
    \end{scope}
    \end{tikzpicture}
    

%% file: figures/figure-pipeline.tex
\begin{tikzpicture}
    \begin{scope}
        \begin{scope}[shift={(0, 0)}]
            \begin{scope}[shift={(0, 1.0)}]
                \draw[rounded corners = 1.5, fill = mygreenlight, anchor = center, line width = 0.5] 
                    (-1.75, -0.25) rectangle (1.75, 0.25);
                \node[black, draw = none, anchor = west] 
                    at (-1.65, 0) {\scalebox{0.75}{allocation counter}};
                \node[black, draw = none, anchor = center] 
                    at (1.25, 0) {\scalebox{0.75}{ \texttt{[RW]} }};
            \end{scope}
            \begin{scope}[shift={(0, 0.5)}]
                \draw[rounded corners = 1.5, fill = mygreylighter, anchor = center, line width = 0.5] 
                    (-1.75, -0.25) rectangle (1.75, 0.25);
                \node[black, draw = none, anchor = center] 
                    at (0, 0) {\scalebox{0.75}{\texttt{Dispatch(1)}}};
            \end{scope}
            \begin{scope}
                \draw[rounded corners = 1.5, fill = myredlight, anchor = center, line width = 0.5] 
                    (-1.75, -0.25) rectangle (1.75, 0.25);
                \node[black, draw = none, anchor = center] 
                    at (0, 0) {\scalebox{1.0}{(1) \texttt{ResetCounter}}};

                \draw[black, line width = 0.5, ->, >=stealth] 
                    (1.75, 0.0) -- (2.25, 0.0) -- (2.25, 0.5) -- (2.75, 0.5);
            \end{scope}
        \end{scope}
        \begin{scope}[shift={(4.5, 0)}]
            \begin{scope}[shift={(0, 1.5)}]
                \draw[rounded corners = 1.5, fill = mygreenlight, anchor = center, line width = 0.5] 
                    (-1.75, -0.25) rectangle (1.75, 0.25);
                \node[black, draw = none, anchor = west] 
                    at (-1.65, 0) {\scalebox{0.75}{CBT}};
                \node[black, draw = none, anchor = center] 
                    at (1.25, 0) {\scalebox{0.75}{ \texttt{[RO]} }};
            \end{scope}
            \begin{scope}[shift={(0, 1.0)}]
                \draw[rounded corners = 1.5, fill = mygreenlight, anchor = center, line width = 0.5] 
                    (-1.75, -0.25) rectangle (1.75, 0.25);
                \node[black, draw = none, anchor = west] 
                    at (-1.65, 0) {\scalebox{0.75}{pointer buffer}};
                \node[black, draw = none, anchor = center] 
                    at (1.25, 0) {\scalebox{0.75}{ \texttt{[RW]} }};
            \end{scope}
            \begin{scope}[shift={(0, 0.5)}]
                \draw[rounded corners = 1.5, fill = mygreylighter, anchor = center, line width = 0.5] 
                    (-1.75, -0.25) rectangle (1.75, 0.25);
                \node[black, draw = none, anchor = center] 
                    at (0, 0) {\scalebox{0.75}{\texttt{DispatchIndirect()}}};
            \end{scope}
            \begin{scope}
                \draw[rounded corners = 1.5, fill = myredlight, anchor = center, line width = 0.5] 
                    (-1.75, -0.25) rectangle (1.75, 0.25);
                \node[black, draw = none, anchor = center] 
                    at (0, 0) {\scalebox{1.0}{(2) \texttt{CachePointers}}};

                \draw[black, line width = 0.5, ->, >=stealth] 
                    (1.75, 0.0) -- (2.25, 0.0) -- (2.25, 0.5) -- (2.75, 0.5);
            \end{scope}
        \end{scope}
        \begin{scope}[shift={(9, 0)}]
            \begin{scope}[shift={(0, 1.5)}]
                \draw[rounded corners = 1.5, fill = mygreenlight, anchor = center, line width = 0.5] 
                    (-1.75, -0.25) rectangle (1.75, 0.25);
                \node[black, draw = none, anchor = west] 
                    at (-1.65, 0) {\scalebox{0.75}{pointer buffer}};
                \node[black, draw = none, anchor = center] 
                    at (1.25, 0) {\scalebox{0.75}{ \texttt{[RO]} }};
            \end{scope}
            \begin{scope}[shift={(0, 1.0)}]
                \draw[rounded corners = 1.5, fill = mygreenlight, anchor = center, line width = 0.5] 
                    (-1.75, -0.25) rectangle (1.75, 0.25);
                \node[black, draw = none, anchor = west] 
                    at (-1.65, 0) {\scalebox{0.75}{bisector pool}};
                \node[black, draw = none, anchor = center] 
                    at (1.25, 0) {\scalebox{0.75}{ \texttt{[RW]} }};
            \end{scope}
            \begin{scope}[shift={(0, 0.5)}]
                \draw[rounded corners = 1.5, fill = mygreylighter, anchor = center, line width = 0.5] 
                    (-1.75, -0.25) rectangle (1.75, 0.25);
                \node[black, draw = none, anchor = center] 
                    at (0, 0) {\scalebox{0.75}{\texttt{DispatchIndirect()}}};
            \end{scope}
            \begin{scope}
                \draw[rounded corners = 1.5, fill = myredlight, anchor = center, line width = 0.5] 
                    (-1.75, -0.25) rectangle (1.75, 0.25);
                \node[black, draw = none, anchor = center] 
                    at (0, 0) {\scalebox{1.0}{(3) \texttt{ResetCommands}}};

                \draw[black, line width = 0.5, ->, >=stealth] 
                    (1.75, 0.0) -- (2.0, 0.0) -- (2.0, -0.5) -- (-11.25, -0.5) -- (-11.25, -3) -- (-10.75, -3);
            \end{scope}
        \end{scope}
    \end{scope}

    \begin{scope}[shift={(0, -3)}]
        \begin{scope}[shift={(0, 0)}]
            \begin{scope}[shift={(0, 2.0)}]
                \draw[rounded corners = 1.5, fill = mygreenlight, anchor = center, line width = 0.5] 
                    (-1.75, -0.25) rectangle (1.75, 0.25);
                \node[black, draw = none, anchor = west] 
                    at (-1.65, 0) {\scalebox{0.75}{pointer buffer}};
                \node[black, draw = none, anchor = center] 
                    at (1.25, 0) {\scalebox{0.75}{ \texttt{[RO]} }};
            \end{scope}
            \begin{scope}[shift={(0, 1.5)}]
                \draw[rounded corners = 1.5, fill = mygreenlight, anchor = center, line width = 0.5] 
                    (-1.75, -0.25) rectangle (1.75, 0.25);
                \node[black, draw = none, anchor = west] 
                    at (-1.65, 0) {\scalebox{0.75}{bisector pool}};
                \node[black, draw = none, anchor = center] 
                    at (1.25, 0) {\scalebox{0.75}{ \texttt{[RW]} }};
            \end{scope}
            \begin{scope}[shift={(0, 1.0)}]
                \draw[rounded corners = 1.5, fill = mygreenlight, anchor = center, line width = 0.5] 
                    (-1.75, -0.25) rectangle (1.75, 0.25);
                \node[black, draw = none, anchor = west] 
                    at (-1.65, 0) {\scalebox{0.75}{allocation counter}};
                \node[black, draw = none, anchor = center] 
                    at (1.25, 0) {\scalebox{0.75}{ \texttt{[RW]} }};
            \end{scope}
            \begin{scope}[shift={(0, 0.5)}]
                \draw[rounded corners = 1.5, fill = mygreylighter, anchor = center, line width = 0.5] 
                    (-1.75, -0.25) rectangle (1.75, 0.25);
                \node[black, draw = none, anchor = center] 
                    at (0, 0) {\scalebox{0.75}{\texttt{DispatchIndirect()}}};
            \end{scope}
            \begin{scope}
                \draw[rounded corners = 1.5, fill = myredlight, anchor = center, line width = 0.5] 
                    (-1.75, -0.25) rectangle (1.75, 0.25);
                \node[black, draw = none, anchor = center] 
                    at (0, 0) {\scalebox{1.0}{(4) \texttt{GenerateCommands}}};

                \draw[black, line width = 0.5, ->, >=stealth] 
                    (1.75, 0.0) -- (2.25, 0.0) -- (2.25, 0.5) -- (2.75, 0.5);
            \end{scope}
        \end{scope}
        \begin{scope}[shift={(4.5, 0)}]
            \begin{scope}[shift={(0, 2.0)}]
                \draw[rounded corners = 1.5, fill = mygreenlight, anchor = center, line width = 0.5] 
                    (-1.75, -0.25) rectangle (1.75, 0.25);
                \node[black, draw = none, anchor = west] 
                    at (-1.65, 0) {\scalebox{0.75}{pointer buffer}};
                \node[black, draw = none, anchor = center] 
                    at (1.25, 0) {\scalebox{0.75}{ \texttt{[RO]} }};
            \end{scope}
            \begin{scope}[shift={(0, 1.5)}]
                \draw[rounded corners = 1.5, fill = mygreenlight, anchor = center, line width = 0.5] 
                    (-1.75, -0.25) rectangle (1.75, 0.25);
                \node[black, draw = none, anchor = west] 
                    at (-1.65, 0) {\scalebox{0.75}{bisector pool}};
                \node[black, draw = none, anchor = center] 
                    at (1.25, 0) {\scalebox{0.75}{ \texttt{[RW]} }};
            \end{scope}
            \begin{scope}[shift={(0, 1.0)}]
                \draw[rounded corners = 1.5, fill = mygreenlight, anchor = center, line width = 0.5] 
                    (-1.75, -0.25) rectangle (1.75, 0.25);
                \node[black, draw = none, anchor = west] 
                    at (-1.65, 0) {\scalebox{0.75}{allocation counter}};
                \node[black, draw = none, anchor = center] 
                    at (1.25, 0) {\scalebox{0.75}{ \texttt{[RW]} }};
            \end{scope}
            \begin{scope}[shift={(0, 0.5)}]
                \draw[rounded corners = 1.5, fill = mygreylighter, anchor = center, line width = 0.5] 
                    (-1.75, -0.25) rectangle (1.75, 0.25);
                \node[black, draw = none, anchor = center] 
                    at (0, 0) {\scalebox{0.75}{\texttt{DispatchIndirect()}}};
            \end{scope}
            \begin{scope}
                \draw[rounded corners = 1.5, fill = myredlight, anchor = center, line width = 0.5] 
                    (-1.75, -0.25) rectangle (1.75, 0.25);
                \node[black, draw = none, anchor = center] 
                    at (0, 0) {\scalebox{1.0}{(5) \texttt{ReserveBlocks}}};

                \draw[black, line width = 0.5, ->, >=stealth] 
                    (1.75, 0.0) -- (2.25, 0.0) -- (2.25, 0.5) -- (2.75, 0.5);
            \end{scope}
        \end{scope}
        \begin{scope}[shift={(9, 0)}]
            \begin{scope}[shift={(0, 1.5)}]
                \draw[rounded corners = 1.5, fill = mygreenlight, anchor = center, line width = 0.5] 
                    (-1.75, -0.25) rectangle (1.75, 0.25);
                \node[black, draw = none, anchor = west] 
                    at (-1.65, 0) {\scalebox{0.75}{pointer buffer}};
                \node[black, draw = none, anchor = center] 
                    at (1.25, 0) {\scalebox{0.75}{ \texttt{[RO]} }};
            \end{scope}
            \begin{scope}[shift={(0, 1.0)}]
                \draw[rounded corners = 1.5, fill = mygreenlight, anchor = center, line width = 0.5] 
                    (-1.75, -0.25) rectangle (1.75, 0.25);
                \node[black, draw = none, anchor = west] 
                    at (-1.65, 0) {\scalebox{0.75}{bisector pool}};
                \node[black, draw = none, anchor = center] 
                    at (1.25, 0) {\scalebox{0.75}{ \texttt{[RW]} }};
            \end{scope}
            \begin{scope}[shift={(0, 0.5)}]
                \draw[rounded corners = 1.5, fill = mygreylighter, anchor = center, line width = 0.5] 
                    (-1.75, -0.25) rectangle (1.75, 0.25);
                \node[black, draw = none, anchor = center] 
                    at (0, 0) {\scalebox{0.75}{\texttt{DispatchIndirect()}}};
            \end{scope}
            \begin{scope}
                \draw[rounded corners = 1.5, fill = myredlight, anchor = center, line width = 0.5] 
                    (-1.75, -0.25) rectangle (1.75, 0.25);
                \node[black, draw = none, anchor = center] 
                    at (0, 0) {\scalebox{1.0}{(6) \texttt{FillNewBlocks}}};

                \draw[black, line width = 0.5, ->, >=stealth] 
                    (1.75, 0.0) -- (2.0, 0.0) -- (2.0, -0.5) -- (-11.25, -0.5) -- (-11.25, -3) -- (-10.75, -3);
            \end{scope}
        \end{scope}      
    \end{scope}

    \begin{scope}[shift={(0, -6)}]
        \begin{scope}[shift={(0, 0)}]
            \begin{scope}[shift={(0, 1.5)}]
                \draw[rounded corners = 1.5, fill = mygreenlight, anchor = center, line width = 0.5] 
                    (-1.75, -0.25) rectangle (1.75, 0.25);
                \node[black, draw = none, anchor = west] 
                    at (-1.65, 0) {\scalebox{0.75}{pointer buffer}};
                \node[black, draw = none, anchor = center] 
                    at (1.25, 0) {\scalebox{0.75}{ \texttt{[RO]} }};
            \end{scope}
            \begin{scope}[shift={(0, 1.0)}]
                \draw[rounded corners = 1.5, fill = mygreenlight, anchor = center, line width = 0.5] 
                    (-1.75, -0.25) rectangle (1.75, 0.25);
                \node[black, draw = none, anchor = west] 
                    at (-1.65, 0) {\scalebox{0.75}{bisector pool}};
                \node[black, draw = none, anchor = center] 
                    at (1.25, 0) {\scalebox{0.75}{ \texttt{[RW]} }};
            \end{scope}
            \begin{scope}[shift={(0, 0.5)}]
                \draw[rounded corners = 1.5, fill = mygreylighter, anchor = center, line width = 0.5] 
                    (-1.75, -0.25) rectangle (1.75, 0.25);
                \node[black, draw = none, anchor = center] 
                    at (0, 0) {\scalebox{0.75}{\texttt{DispatchIndirect()}}};
            \end{scope}
            \begin{scope}
                \draw[rounded corners = 1.5, fill = myredlight, anchor = center, line width = 0.5] 
                    (-1.75, -0.25) rectangle (1.75, 0.25);
                \node[black, draw = none, anchor = center] 
                    at (0, 0) {\scalebox{1.0}{(7) \texttt{UpdateNeighbors}}};

                \draw[black, line width = 0.5, ->, >=stealth] 
                    (1.75, 0.0) -- (2.25, 0.0) -- (2.25, 0.5) -- (2.75, 0.5);
            \end{scope}
        \end{scope}
        \begin{scope}[shift={(4.5, 0)}]
            \begin{scope}[shift={(0, 2.0)}]
                \draw[rounded corners = 1.5, fill = mygreenlight, anchor = center, line width = 0.5] 
                    (-1.75, -0.25) rectangle (1.75, 0.25);
                \node[black, draw = none, anchor = west] 
                    at (-1.65, 0) {\scalebox{0.75}{CBT}};
                \node[black, draw = none, anchor = center] 
                    at (1.25, 0) {\scalebox{0.75}{ \texttt{[RW]} }};
            \end{scope}
            \begin{scope}[shift={(0, 1.5)}]
                \draw[rounded corners = 1.5, fill = mygreenlight, anchor = center, line width = 0.5] 
                    (-1.75, -0.25) rectangle (1.75, 0.25);
                \node[black, draw = none, anchor = west] 
                    at (-1.65, 0) {\scalebox{0.75}{pointer buffer}};
                \node[black, draw = none, anchor = center] 
                    at (1.25, 0) {\scalebox{0.75}{ \texttt{[RO]} }};
            \end{scope}
            \begin{scope}[shift={(0, 1.0)}]
                \draw[rounded corners = 1.5, fill = mygreenlight, anchor = center, line width = 0.5] 
                    (-1.75, -0.25) rectangle (1.75, 0.25);
                \node[black, draw = none, anchor = west] 
                    at (-1.65, 0) {\scalebox{0.75}{bisector pool}};
                \node[black, draw = none, anchor = center] 
                    at (1.25, 0) {\scalebox{0.75}{ \texttt{[RW]} }};
            \end{scope}
            \begin{scope}[shift={(0, 0.5)}]
                \draw[rounded corners = 1.5, fill = mygreylighter, anchor = center, line width = 0.5] 
                    (-1.75, -0.25) rectangle (1.75, 0.25);
                \node[black, draw = none, anchor = center] 
                    at (0, 0) {\scalebox{0.75}{\texttt{DispatchIndirect()}}};
            \end{scope}
            \begin{scope}
                \draw[rounded corners = 1.5, fill = myredlight, anchor = center, line width = 0.5] 
                    (-1.75, -0.25) rectangle (1.75, 0.25);
                \node[black, draw = none, anchor = center] 
                    at (0, 0) {\scalebox{1.0}{(8) \texttt{UpdateBitfield}}};

                \draw[black, line width = 0.5, ->, >=stealth] 
                    (1.75, 0.0) -- (2.25, 0.0) -- (2.25, 1.0) -- (2.75, 1.0);
            \end{scope}
        \end{scope}
        \begin{scope}[shift={(9, 0)}]
            \begin{scope}[shift={(0, 1.5)}]
                \draw[rounded corners = 1.5, fill = mygreenlight, anchor = center, line width = 0.5] 
                    (-1.75, -0.25) rectangle (1.75, 0.25);
                \node[black, draw = none, anchor = west] 
                    at (-1.65, 0) {\scalebox{0.7}{CBT}};
                \node[black, draw = none, anchor = center] 
                    at (1.25, 0) {\scalebox{0.75}{ \texttt{[RW]} }};
            \end{scope}
            \begin{scope}[shift={(0, 1)}]
                \draw[rounded corners = 1.5, fill = mygreylighter, anchor = center, line width = 0.5] 
                    (-1.75, -0.25) rectangle (1.75, 0.25);
                \node[black, draw = none, anchor = center] 
                    at (0, 0) {\scalebox{0.75}{\texttt{for each $d$ in [$0,\;D-1$]}}};
            \end{scope}
            \begin{scope}[shift={(0, 0.5)}]
                \draw[rounded corners = 1.5, fill = mygreylighter, anchor = center, line width = 0.5] 
                    (-1.75, -0.25) rectangle (1.75, 0.25);
                \node[black, draw = none, anchor = center] 
                    at (0, 0) {\scalebox{0.75}{\texttt{Dispatch($2^{D-1-d}$)}}};
            \end{scope}
            \begin{scope}
                \draw[rounded corners = 1.5, fill = myredlight, anchor = center, line width = 0.5] 
                    (-1.75, -0.25) rectangle (1.75, 0.25);
                \node[black, draw = none, anchor = center] 
                    at (0, 0) {\scalebox{1.0}{(9) \texttt{SumReduction}}};

                \draw[black, line width = 0.5, ->, >=stealth] 
                    (1.75, 0.0) -- (2.25, 0.0) -- (2.25, 8.0) -- (-11.25, 8.0) -- (-11.25, 6.0) -- (-10.75, 6.0);
            \end{scope}
        \end{scope}      
    \end{scope}

    \end{tikzpicture}

%% file: figures/figure-moon.tex
\begin{center}
\setlength{\tabcolsep}{1.5pt}
\begin{tabular}{c c c c}
    \begin{tikzpicture}[every node/.style={inner sep=0,outer sep=0}]
        \draw (0, 0) node[anchor=north west] {\includegraphics[fbox, width=0.23\columnwidth]{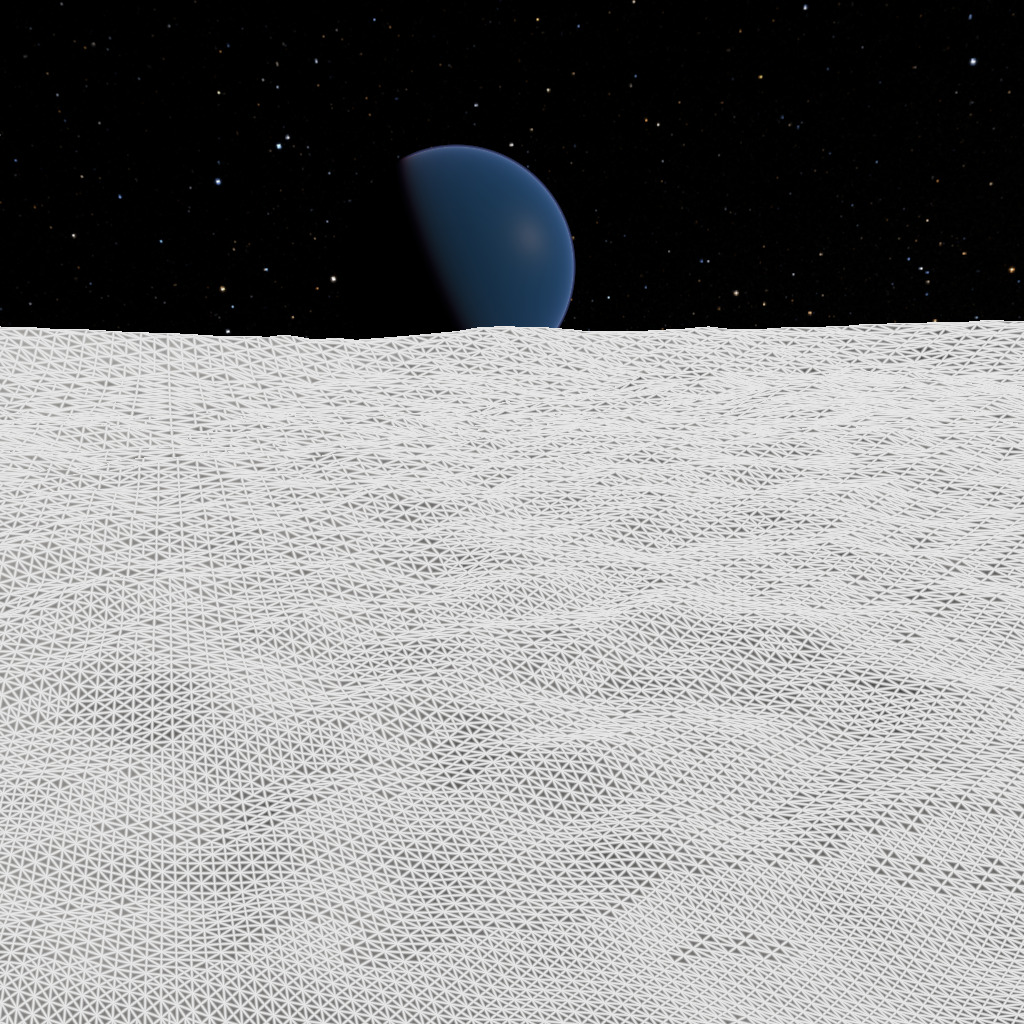}};
    \end{tikzpicture}
    &
    \begin{tikzpicture}[every node/.style={inner sep=0,outer sep=0}]
        \draw (0, 0) node[anchor=north west] {\includegraphics[fbox, width=0.23\columnwidth]{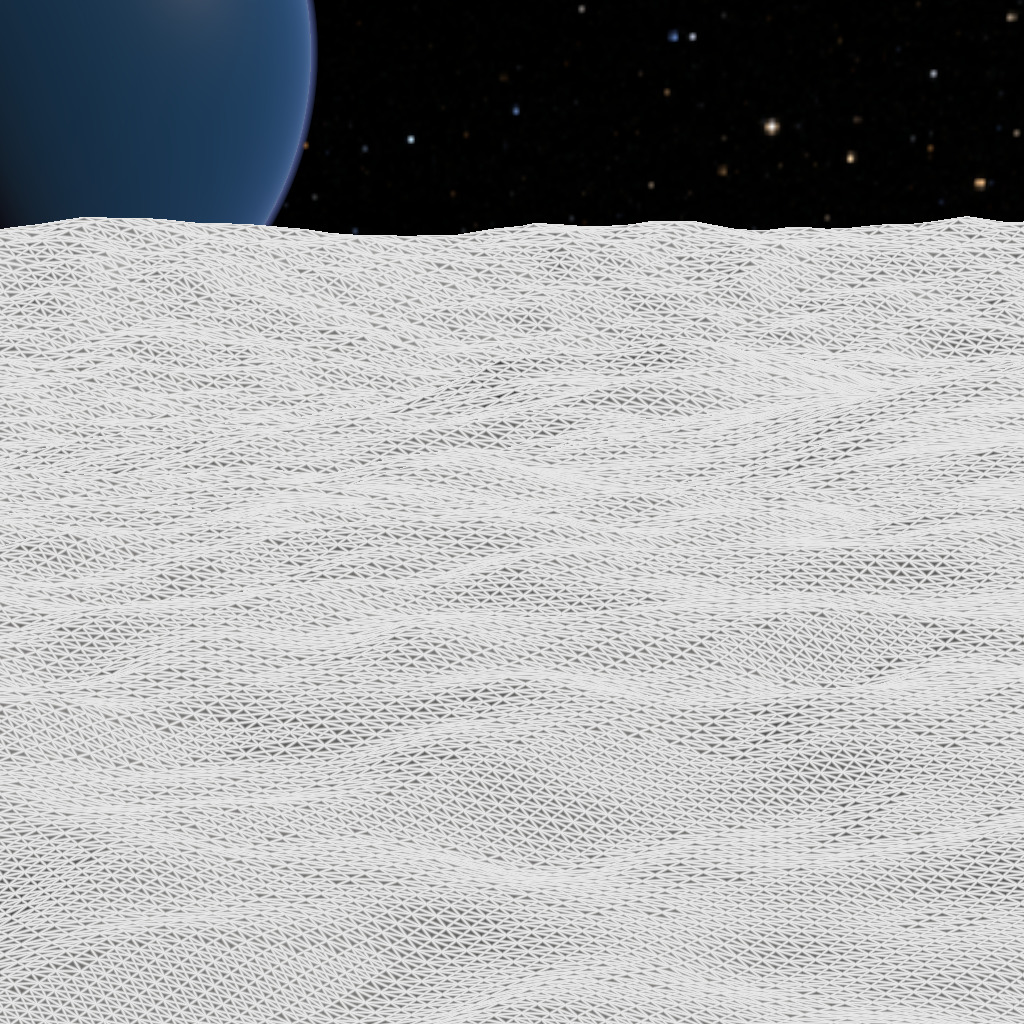}};
    \end{tikzpicture}
    &
    \begin{tikzpicture}[every node/.style={inner sep=0,outer sep=0}]
        \draw (0, 0) node[anchor=north west] {\includegraphics[fbox, width=0.23\columnwidth]{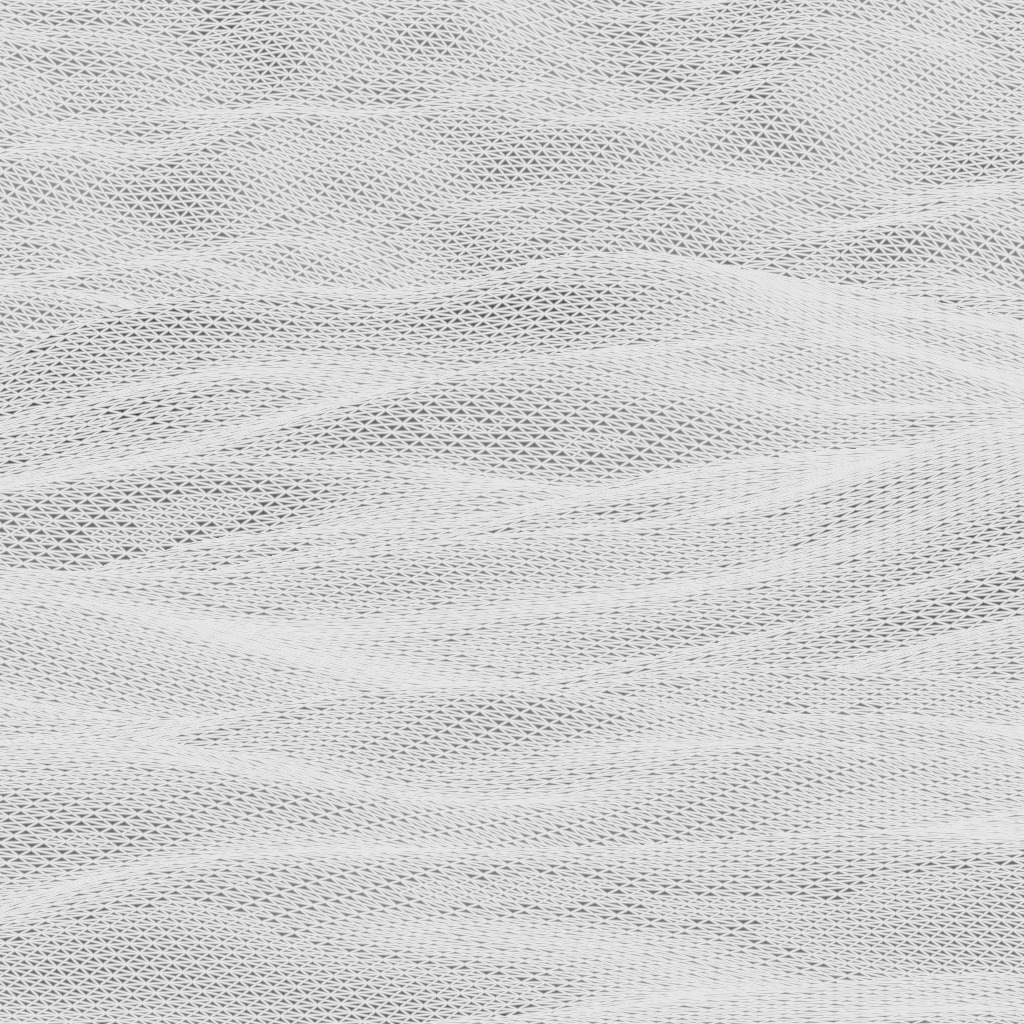}};
    \end{tikzpicture}
    &
    \begin{tikzpicture}[every node/.style={inner sep=0,outer sep=0}]
        \draw (0, 0) node[anchor=north west] {\includegraphics[fbox, width=0.23\columnwidth]{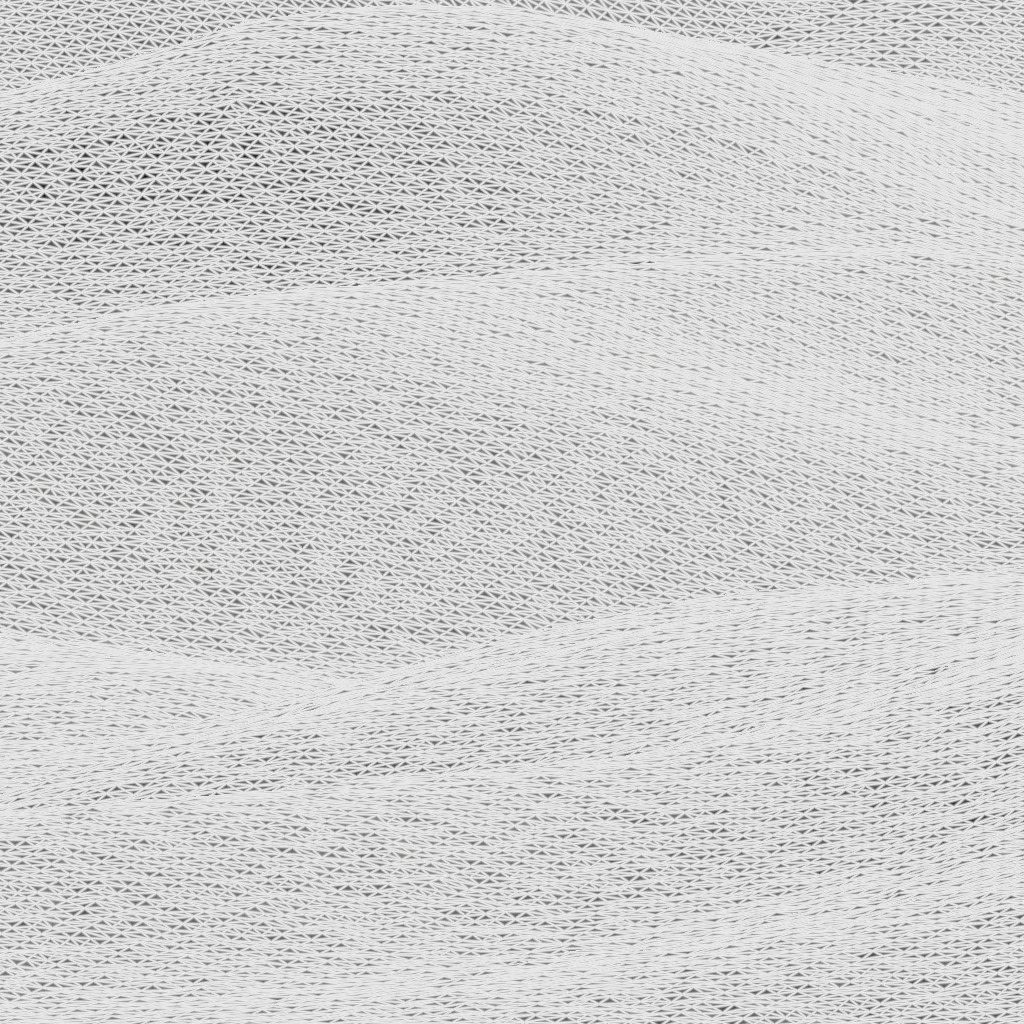}};
    \end{tikzpicture}
    \\
    \begin{tikzpicture}[every node/.style={inner sep=0,outer sep=0}]
        \draw (0, 0) node[anchor=north west] {\includegraphics[fbox, width=0.23\columnwidth]{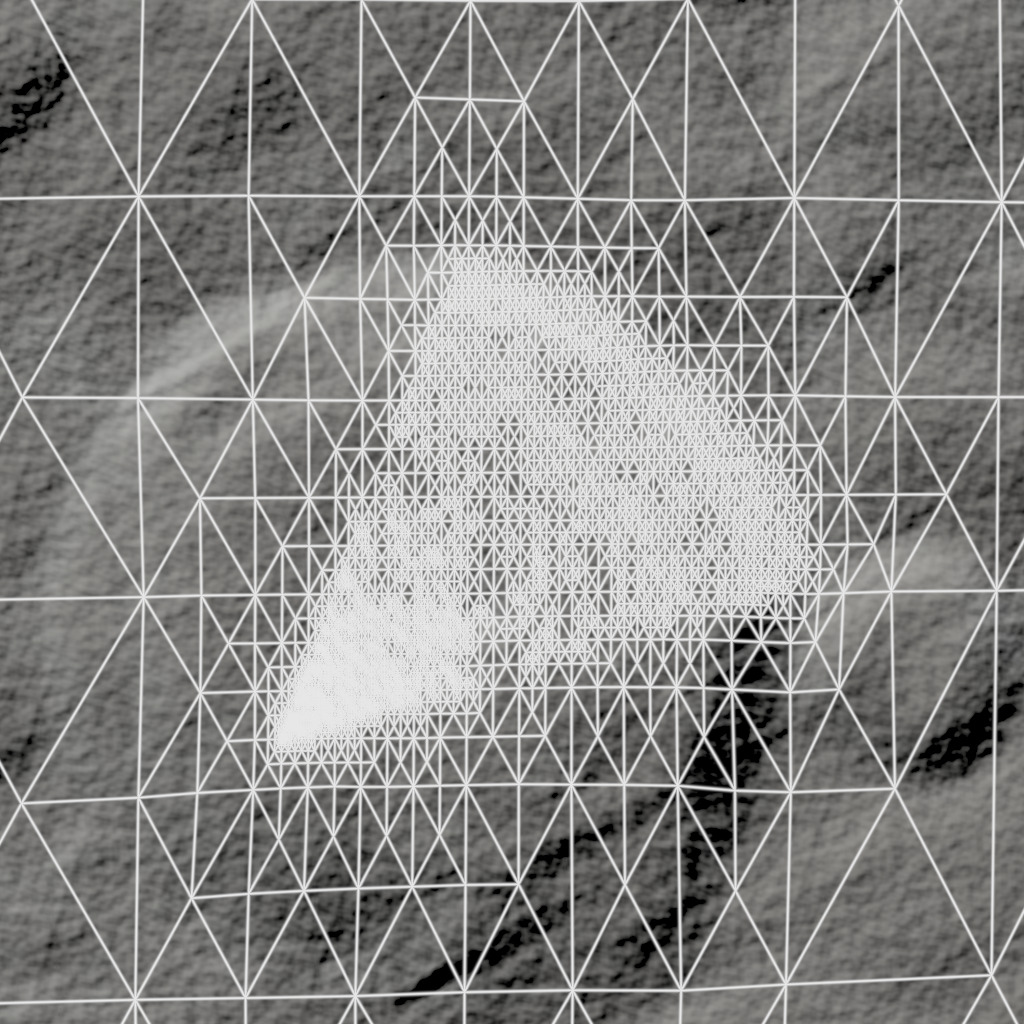}};
    \end{tikzpicture}
    &
    \begin{tikzpicture}[every node/.style={inner sep=0,outer sep=0}]
        \draw (0, 0) node[anchor=north west] {\includegraphics[fbox, width=0.23\columnwidth]{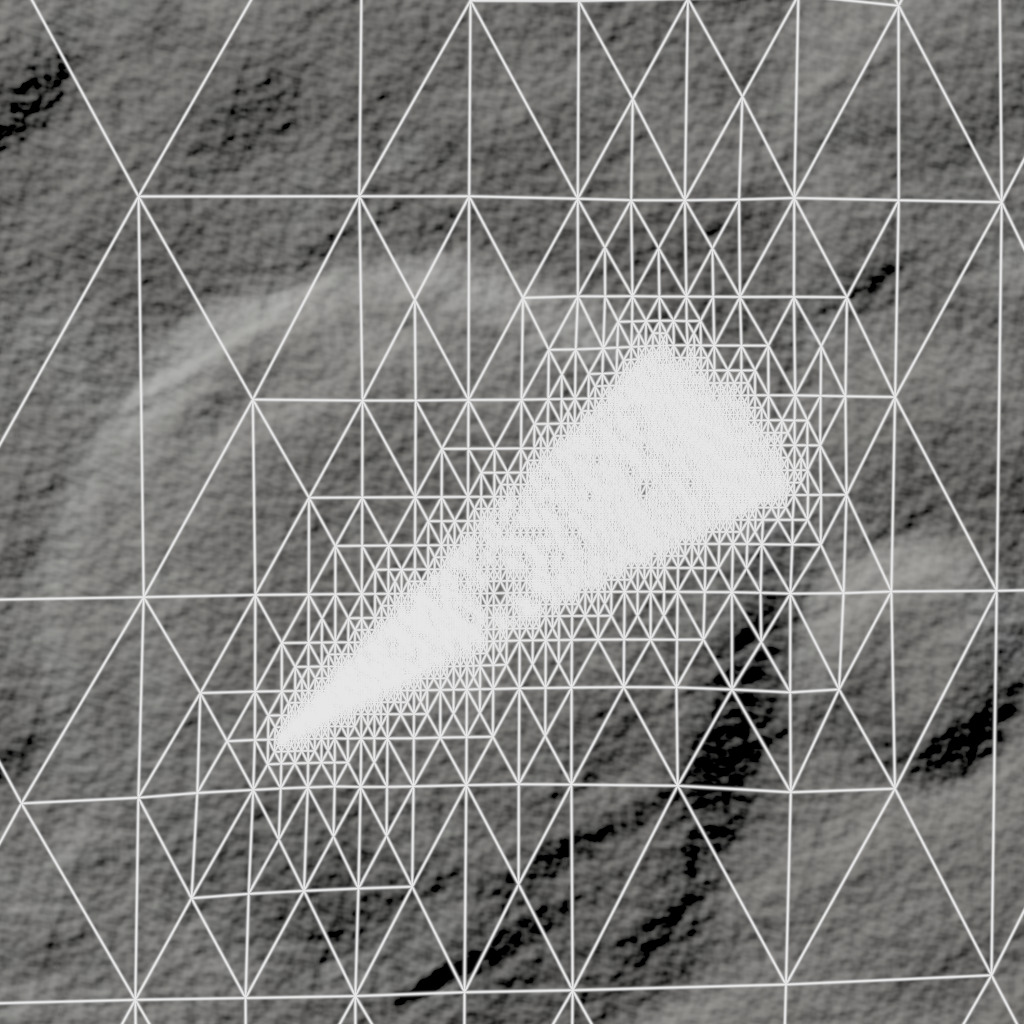}};
    \end{tikzpicture}
    &
    \begin{tikzpicture}[every node/.style={inner sep=0,outer sep=0}]
        \draw (0, 0) node[anchor=north west] {\includegraphics[fbox, width=0.23\columnwidth]{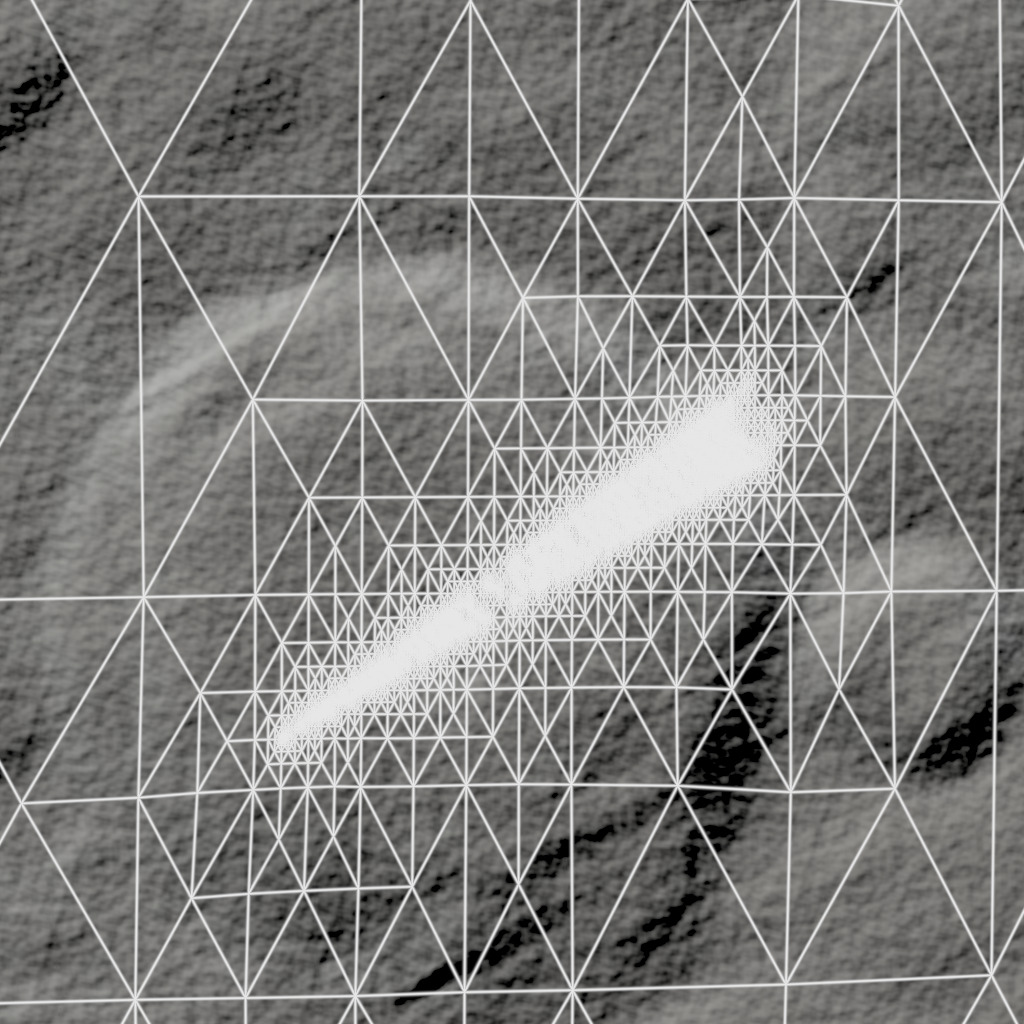}};
    \end{tikzpicture}
    &
    \begin{tikzpicture}[every node/.style={inner sep=0,outer sep=0}]
        \draw (0, 0) node[anchor=north west] {\includegraphics[fbox, width=0.23\columnwidth]{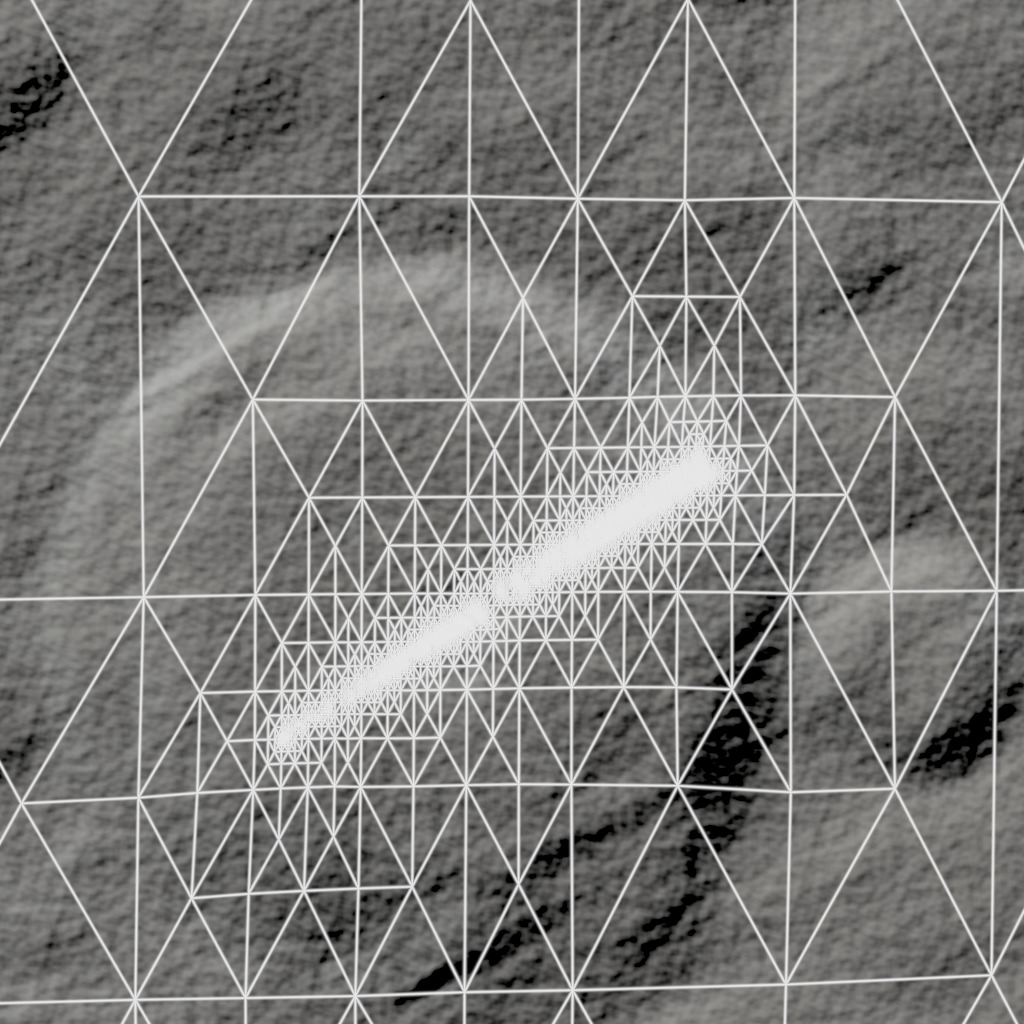}};
    \end{tikzpicture}
\end{tabular}
\end{center}   

%% file: figures/figure-tessellation.tex
\setlength{\tabcolsep}{1.5pt}
\begin{center}
\begin{tabular}{c c}
    \begin{tikzpicture}[every node/.style={inner sep=0,outer sep=0}]
        \draw (0, 0) node[anchor=north west] {\includegraphics[fbox, width=0.47\columnwidth]{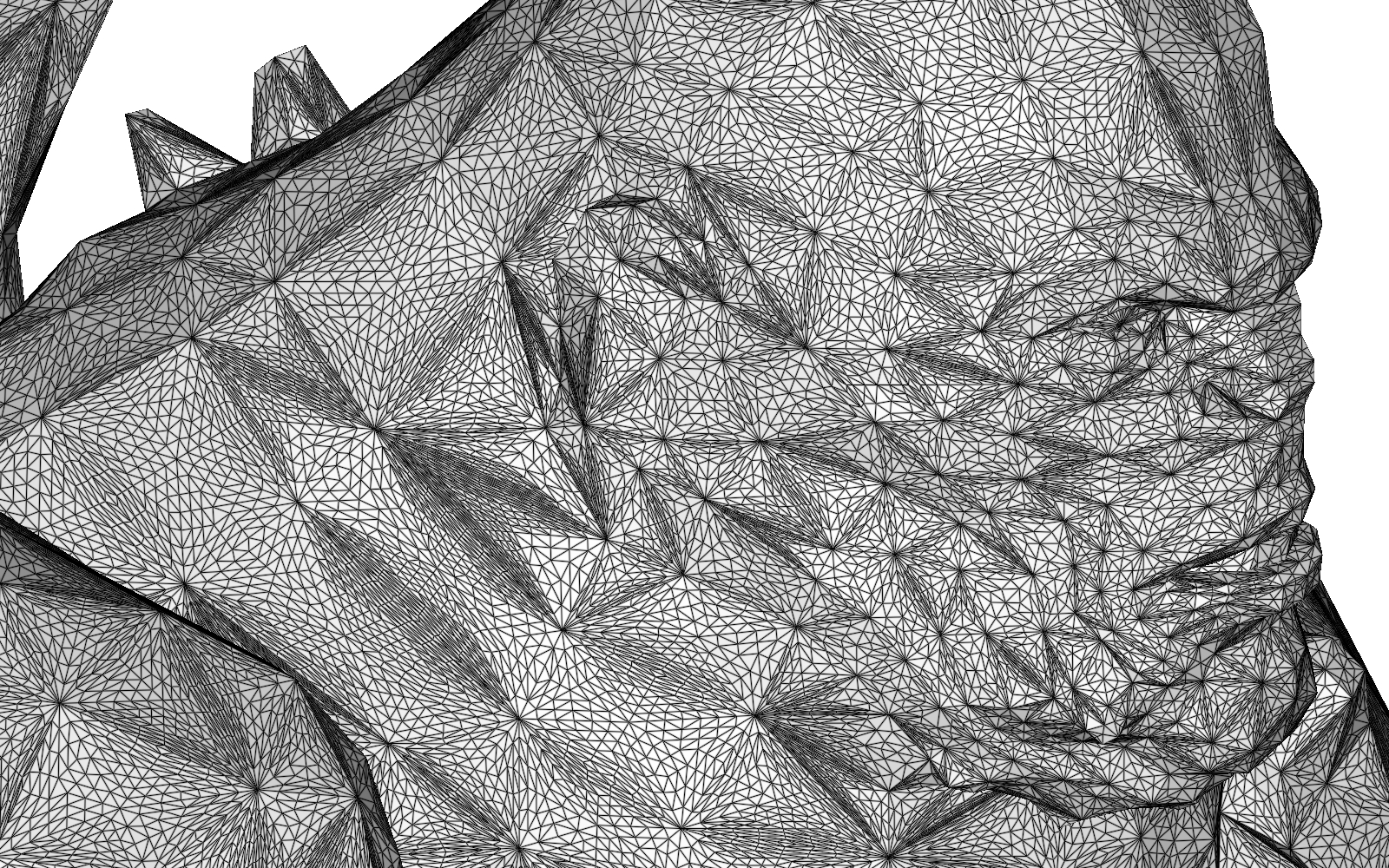}};
    \end{tikzpicture}
    &
    \begin{tikzpicture}[every node/.style={inner sep=0,outer sep=0}]
        \draw (0, 0) node[anchor=north west] {\includegraphics[fbox, width=0.47\columnwidth]{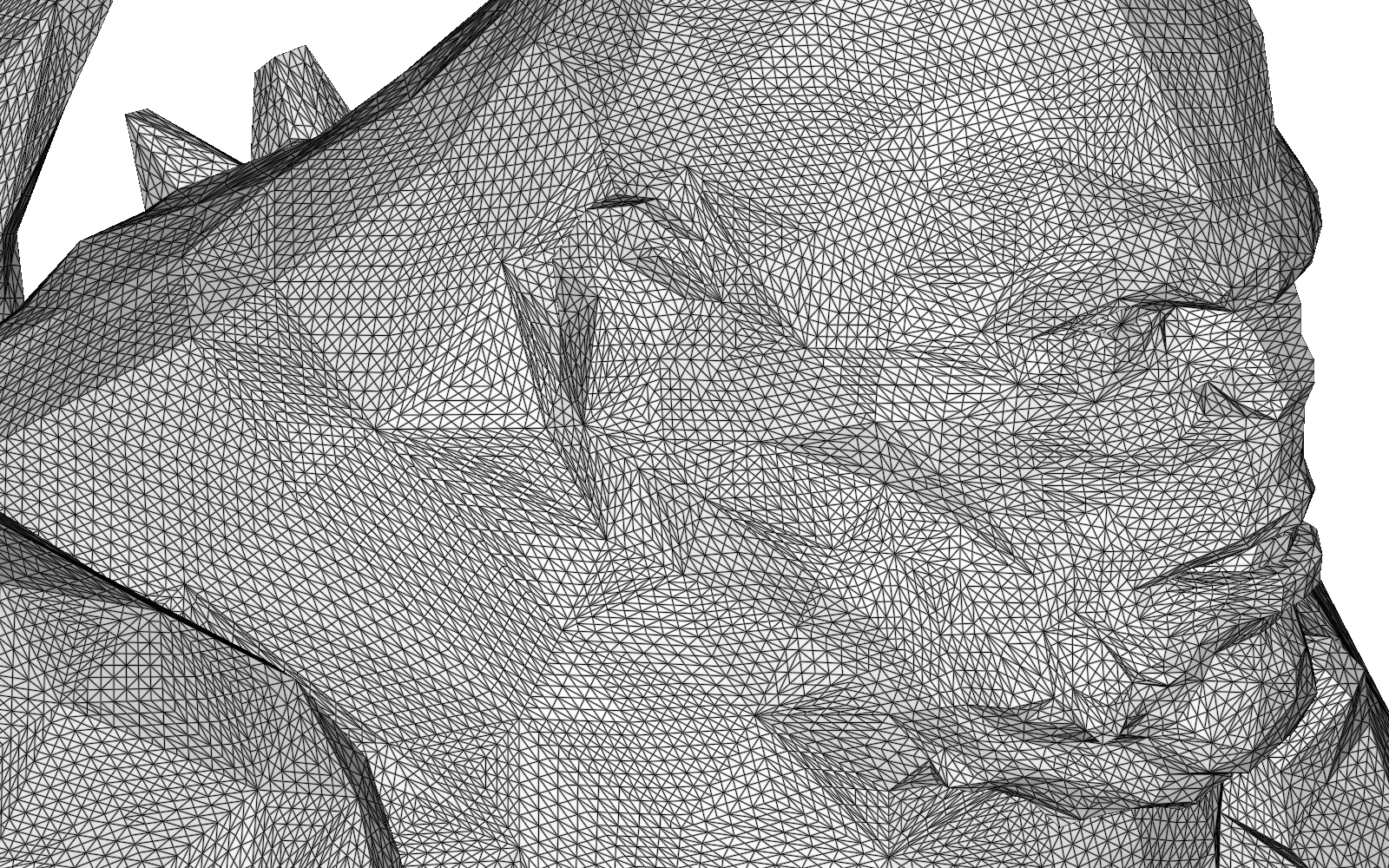}};
    \end{tikzpicture}
    \\
    \begin{tikzpicture}[every node/.style={inner sep=0,outer sep=0}]
        \draw (0, 0) node[anchor=north west] {\includegraphics[fbox, width=0.47\columnwidth]{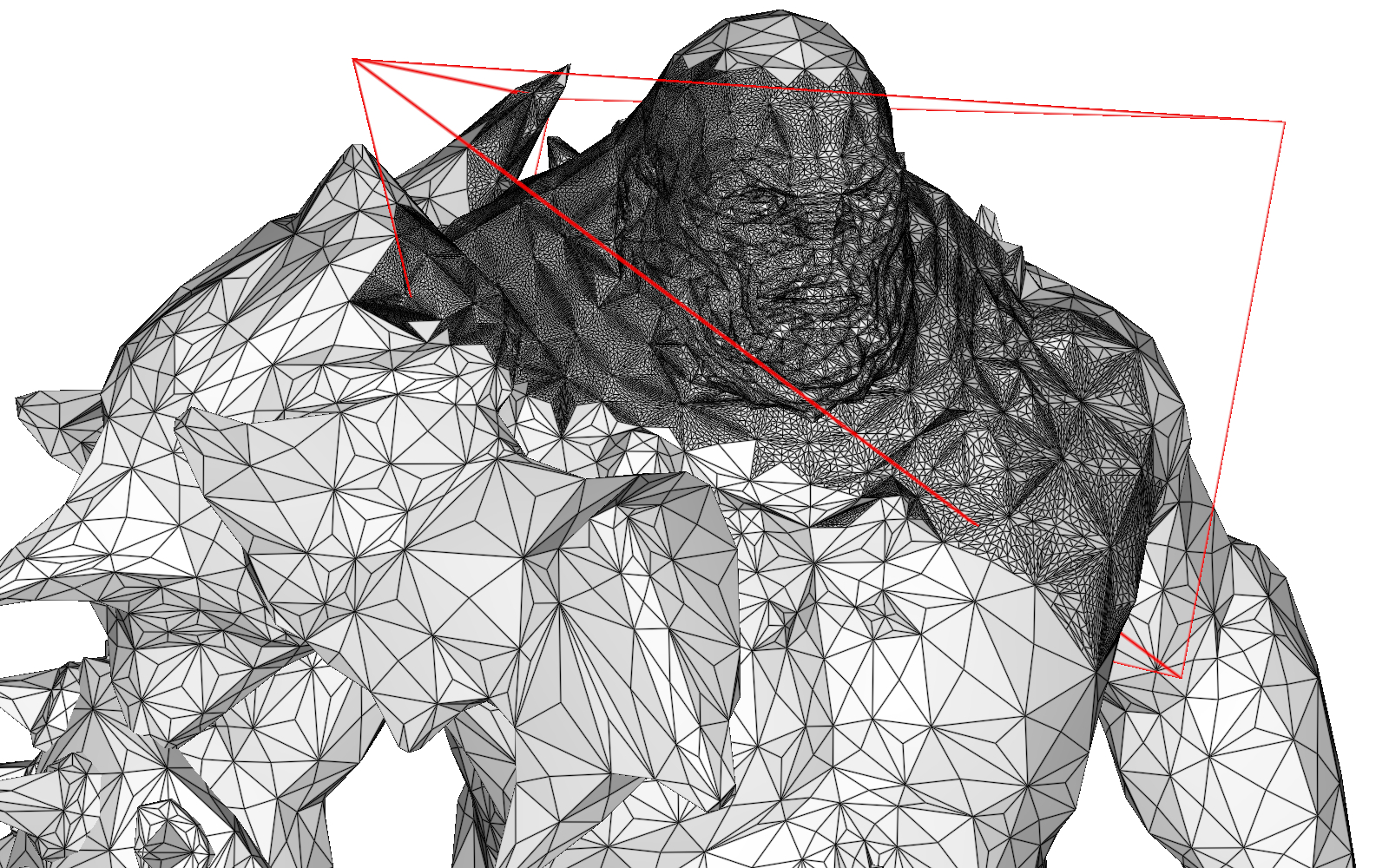}};
    \end{tikzpicture}
    &
    \begin{tikzpicture}[every node/.style={inner sep=0,outer sep=0}]
        \draw (0, 0) node[anchor=north west] {\includegraphics[fbox, width=0.47\columnwidth]{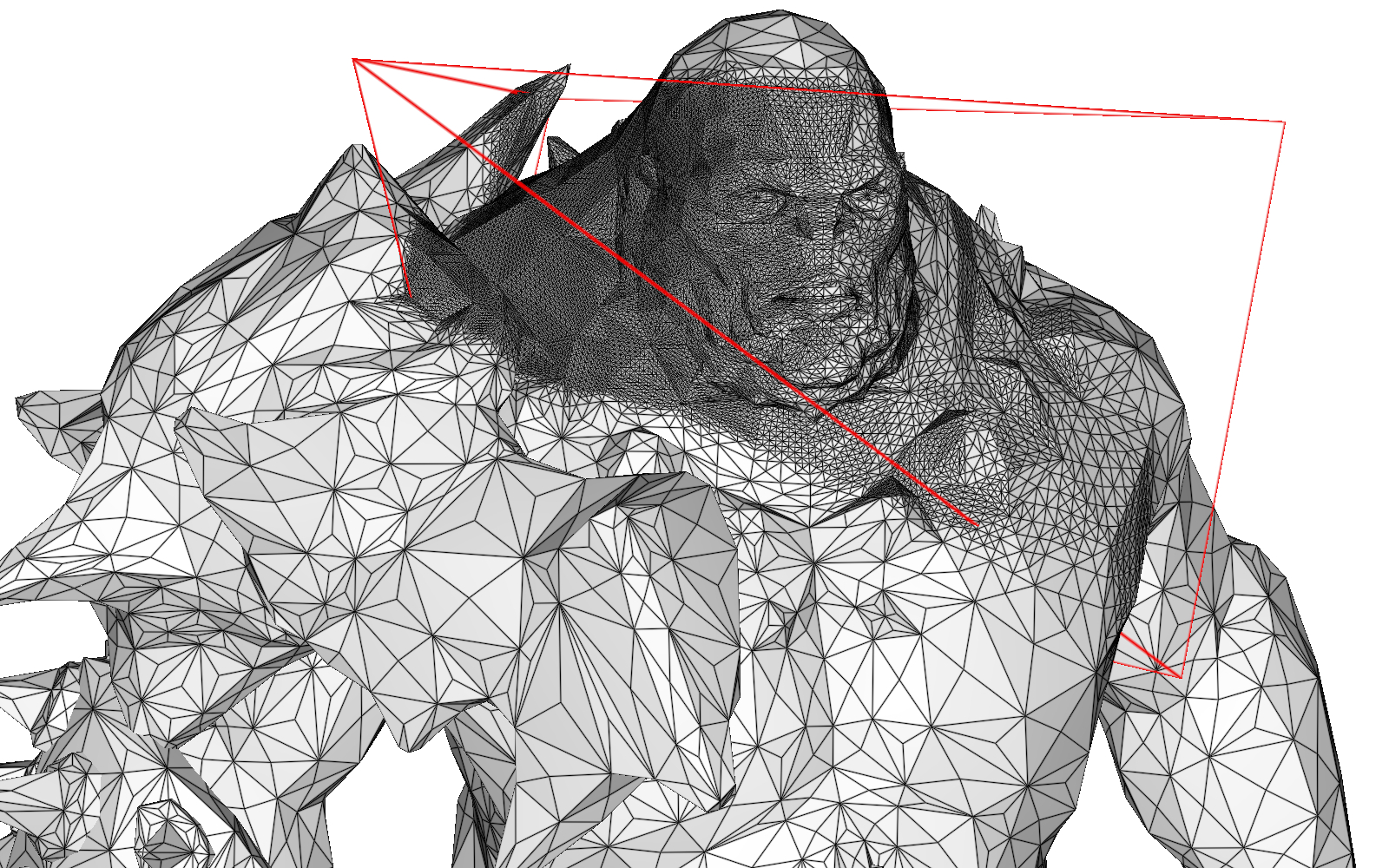}};
    \end{tikzpicture}
\end{tabular}
\end{center}   

%% file: figures/figure-float.tex
\setlength{\tabcolsep}{1.5pt}
\begin{center}
\begin{tabular}{c c}
    \begin{tikzpicture}[every node/.style={inner sep=0,outer sep=0}]
        \draw (0, 0) node[anchor=north west] {\includegraphics[fbox, width=0.47\columnwidth]{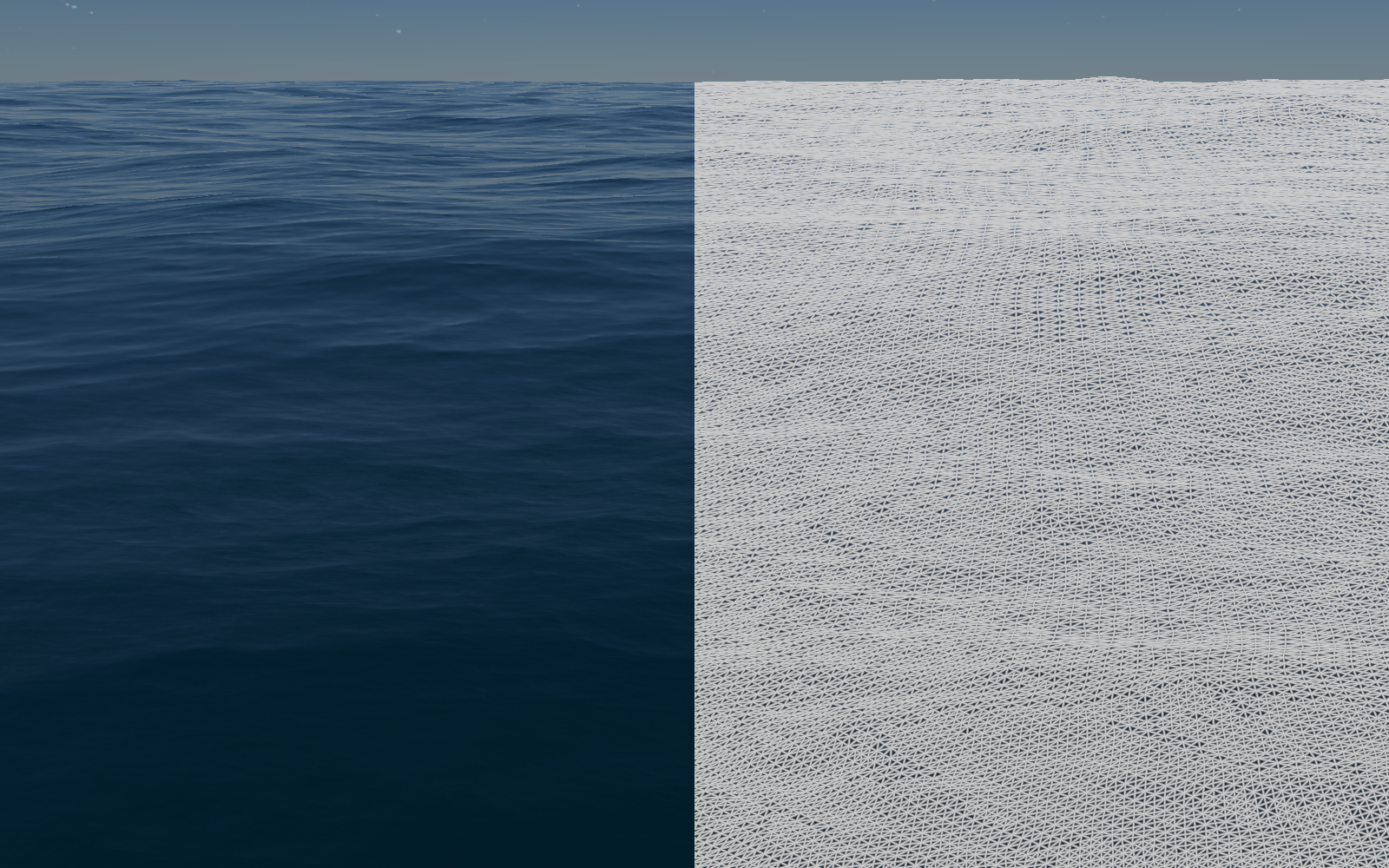}};
        \node [mygreylighter, anchor = north west] at (0.05, -0.05) {{[fp64]}};
        \draw[myred, line width = 0.5] (3.27, -4.08) -- (3.27, -0.01);
    \end{tikzpicture}
    &
    \begin{tikzpicture}[every node/.style={inner sep=0,outer sep=0}]
        \draw (0, 0) node[anchor=north west] {\includegraphics[fbox, width=0.47\columnwidth]{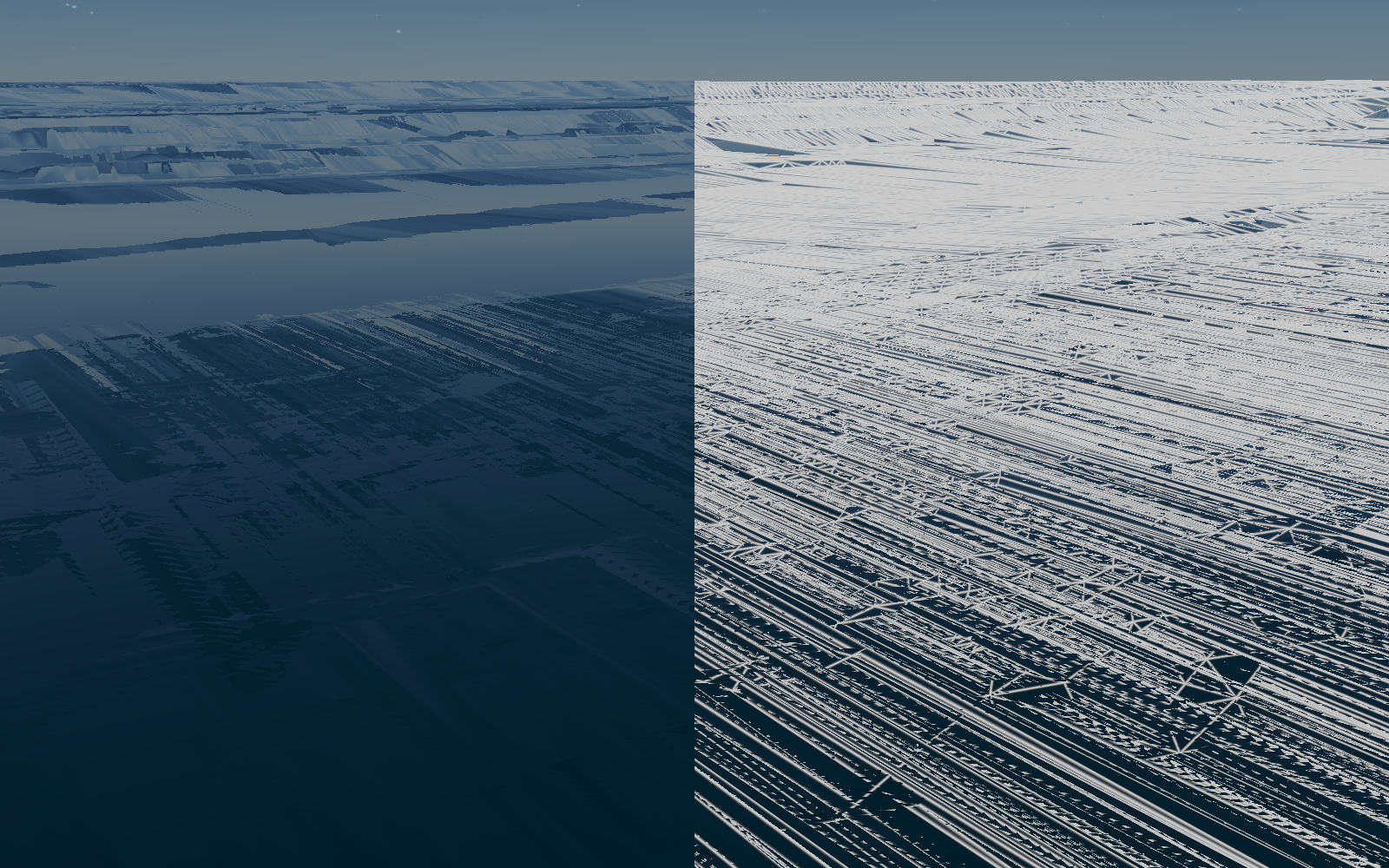}};
        \node [mygreylighter, anchor = north west] at (0.05, -0.05) {{[fp32]}};
        \draw[myred, line width = 0.5] (3.27, -4.08) -- (3.27, -0.01);
    \end{tikzpicture}
\end{tabular}
\end{center}   